\documentclass[12pt]{article} 
\usepackage{graphicx,subfigure,latexsym}
\usepackage{amsmath,amssymb,amsfonts}
\usepackage{bm}

\newcommand{\be}{\begin{equation}}
\newcommand{\ee}{\end{equation}}
\newcommand{\bear}{\begin{eqnarray}}
\newcommand{\eear}{\end{eqnarray}}
\newcommand{\ba}{\begin{array}}
\newcommand{\ea}{\end{array}}

\def \be {\begin{equation}}
\def \ee {\end{equation}}

\def \bes {\begin{subequations}}
\def \ees {\end{subequations}}

\def \<{\langle}
\def \>{\rangle}
\def \+{\dagger}
\def \({\left(}
\def \){\right)}
\def \[{\left[}
\def \]{\right]}

\def \m {\mu}

\begin{document}

\begin{titlepage}
\vfill
\begin{flushright}
{\normalsize RBRC-1072}\\
\end{flushright}

\vfill
\begin{center}
{\Large\bf  Chiral Magnetic and Vortical Effects \\in Higher Dimensions at Weak Coupling}

\vskip 0.3in

\vskip 0.3in
Ho-Ung Yee\footnote{e-mail:
{\tt hyee@uic.edu}}
\vskip 0.15in

{\it  Department of Physics, University of Illinois, Chicago, Illinois 60607}\\{\it and}\\
{\it  RIKEN-BNL Research Center, Brookhaven National Laboratory,}\\
{\it Upton, New York
11973-5000}
\\[0.15in]
{\normalsize  2014}

\end{center}

\vfill

\begin{abstract}

Chiral Magnetic Effect (CME) and Chiral Vortical Effect (CVE) are parity odd transport phenomena originating from chiral anomaly, and have generalizations to
all even dimensional space-time higher than four dimensions. We attempt to compute the associated P-odd retarded response functions in the weak coupling limit of chiral fermion theory in all even dimensions, using the diagrammatic technique of real-time perturbation theory. We also clarify the necessary Kubo formula relating the computed P-odd retarded correlation functions and the associated anomalous transport coefficients.
We speculate on the 8-fold classification of topological phases.
\end{abstract}

\vfill

\end{titlepage}
\setcounter{footnote}{0}

\baselineskip 18pt \pagebreak
\renewcommand{\thepage}{\arabic{page}}
\pagebreak

\section{Introduction}

The physics of chiral anomaly in four space-time dimensions has been explored extensively, which leads to many interesting dynamical phenomena, while at the same time, many of them are topologically protected against possible modifications due to interactions.
Hydrodynamic transport phenomena arising from chiral anomaly in the finite temperature/density regime
have received a recent surge of interest, partly due to their importance in heavy-ion collisions and condensed matter systems of Weyl semimetals. At leading order in derivative expansion, there exist Chiral Magnetic Effect (CME) \cite{Kharzeev:2007tn,Kharzeev:2007jp,Fukushima:2008xe,Son:2004tq,Metlitski:2005pr} and Chiral Vortical Effect (CVE) \cite{Erdmenger:2008rm,Banerjee:2008th}.
The CME is the phenomenon of induced current along the direction of the applied magnetic field,
\be
\vec J=\sigma_\chi \vec B\,,
\ee
with a chiral magnetic conductivity $\sigma_\chi$. For the system of a single Weyl fermion in four dimensions with a chemical potential $\mu$, we have 
\be
\sigma_\chi={\mu\over 4\pi^2}\,.
\ee
For the CVE, the fluid vorticity $\vec\omega=(1/2)\vec\nabla\times\vec v$ plays a role of magnetic field instead,
\be
\vec J=\sigma_V \vec\omega\,,
\ee
with the chiral vortical conductivity for a single Weyl spinor
\be
\sigma_V={1\over 4\pi^2}\left(\mu^2+{\pi^2\over 3} T^2\right)\,.
\ee
In addition to the above anomaly induced charge current, there also appears anomaly induced energy flow, or momentum density, $T^{0i}\equiv \vec P$ \cite{Loganayagam:2011mu,Landsteiner:2012kd,Yee:2009vw}. For a single Weyl fermion, we have
\be
\vec P=\left({1\over 8\pi^2}\mu^2+{1\over 24}T^2\right)\vec B+\left({1\over 6\pi^2}\mu^3+{1\over 6}\mu T^2\right)\vec\omega\,.
\ee
Interestingly, these anomaly induced transport coefficients can be fixed by a purely hydrodynamic consideration of the second law of thermodynamics \cite{Son:2009tf}, that is, the non-decrease of entropy in time, except 
the pieces in the above containing $T^2$ which have been argued to be related to the mixed current-gravitational anomaly \cite{Landsteiner:2011cp}.
However, there also exist different claims on the origin of such $T^2$ corrections, for example, Ref.\cite{Basar:2013qia,Braguta:2014gea,Kalaydzhyan:2014bfa}. The values we show in the above are from the free fermion computations \cite{Landsteiner:2011cp,Kharzeev:2009pj,Loganayagam:2012pz}, and there are some demonstrations of their universality in strong coupling holography \cite{Landsteiner:2011iq}, in a perturbative weak coupling Yukawa theory \cite{Golkar:2012kb}, and in effective action approach \cite{Jensen:2012kj,Banerjee:2012cr,Jensen:2013kka,Jensen:2013rga,Haehl:2013hoa}.

The CME and CVE have generalizations in even space-time dimensions higher than four \cite{Loganayagam:2011mu,Kharzeev:2011ds}. Instead of magnetic field or vorticity, we have a set of several P-odd vectors: in $2n$ dimensions there are $n$ possible such vectors as
\be
B^\mu_{(s,t)}\equiv {1\over n}\epsilon^{\mu\nu\mu_1\nu_1\cdots\mu_{n-1}\nu_{n-1}}u_\nu\left(\partial_{\mu_1}u_{\nu_1}\right)\cdots\left(\partial_{\mu_s} u_{\nu_s}\right) F_{\mu_{s+1}\nu_{s+1}}\cdots F_{\mu_{n-1}\nu_{n-1}}\,,\label{Bst0}
\ee
where $s$ runs from $0$ to $(n-1)$ with $s+t=(n-1)$, and the generalized CME/CVE is
\be
J^\mu=\sum_{s=0}^{n-1} \xi_{(s,t)} B^\mu_{(s,t)}\,,\quad T^{0\mu}=\sum_{s=0}^{n-1} \lambda_{(s,t)} B^\mu_{(s,t)}\,,
\ee
with a set of $2n$ transport coefficients $\xi_{(s,t)}$ and $\lambda_{(s,t)}$ \footnote{
In the Landau frame, one has to redefine the fluid velocity such that $\lambda_{(s,t)}=0$, which in turn shifts the value of $\xi_{(s,t)}$. See our discussion near the end of Section 5 on this frame choice issue.}.
In Refs.\cite{Loganayagam:2011mu,Kharzeev:2011ds}, these coefficients, up to polynomials of temperature like $T^2$ in four dimensions, have been analytically determined 
in the hydrodynamic framework by requiring the principle of time-reversal invariance or non-generation of entropy by these transport terms.
Ref.\cite{Loganayagam:2012pz} takes a further microscopic view on this principle in the free fermion limit based on the notion of topologically protected chiral zero modes to derive full expressions for $\xi_{(s,t)}$ and $\lambda_{(s,t)}$ including temperature corrections.

The purpose of this work is to provide an explicit diagrammatic computation of $\xi_{(s,t)}$ and $\lambda_{(s,t)}$ in free chiral fermion theory, with the clarification on the relevant Kubo formula connecting the P-odd retarded correlation functions of current and energy-momentum operators to the transport coefficients $\xi_{(s,t)}$ and $\lambda_{(s,t)}$. 
The first P-odd retarded response functions appear at $(n-1)$'th order of the external gauge and metric perturbations.
We will also clarify the subtleties regarding the frame choice, which might be a useful addition to the existing literature, too.

Our computation leads to two integral identities, (\ref{conj1}) and (\ref{conj2}), which we couldn't prove, but have been checked
explicitly for some low $n$ values. With these two mathematical identities accepted, we are able to sum up all the diagrams with many different topologies analytically in real-time perturbation theory for the first non-trivial P-odd contributions at zero frequency-momentum limit. The resulting values of $\xi_{(s,t)}$ and $\lambda_{(s,t)}$ from these P-odd retarded correlation functions after using the developed Kubo formula agree remarkably with the hydrodynamic predictions.
Since the summation of many different diagrams is quite non-trivial and intricate involving several combinatoric identities, this agreement is a convincing retrospective evidence for our two conjectured mathematical identities.

\section{Basics of chiral spinors in $d+1=2n$ dimensions}

This section serves as a summary of the relevant facts about the chiral spinors in the general even dimensions $d+1=2n$ that we are going to use in the following sections ($d$ denotes the number of space dimensions). It will also fix our notations and conventions.

We start from a massless Dirac spinor in $d+1=2n$ which consists of a pair of chiral spinors with different chirality. We will eventually pick only one chiral spinor out of this Dirac spinor.
The Dirac action reads as
\be
{\cal L}=\bar\psi \gamma^\mu \left(\partial_\mu-ieA_\mu\right)\psi\,,
\ee
where our metric convention is $\eta={\rm diag}(-,+,\cdots,+)$ (mostly positive convention), and
\be
\bar\psi\equiv -i\psi^\dagger \gamma^0\,.
\ee
The Dirac matrices satisfy the usual relation
\be
\{\gamma^\mu,\gamma^\nu\}=2\eta^{\mu\nu}\,,\label{anti}
\ee
so that $\gamma^0$ is anti-hermitian in our convention. The Dirac matrices are $2^n\times 2^n$ matrices.
Upon quantization, the spinor operators satisfy the equal-time commutation relation
\be
\{\psi^\dagger_\alpha (\vec x),\psi_\beta(\vec y)\}=\delta^{(d)}(\vec x-\vec y)\delta_{\alpha\beta}\,,
\ee
where $\alpha,\beta$ run over $2^n$-components of the spinor index.

To perform a projection to one chiral component of $2^{n-1}$ dimensions, we define $\gamma^5$ as
\be
\gamma^5\equiv i^{n-1}\gamma^0\gamma^1\cdots\gamma^{2n-1}\,,
\ee
which anti-commutes with all $\gamma^\mu$'s and satisfies
\be
(\gamma^5)^2={\bf 1}\,,\quad (\gamma^5)^\dagger =\gamma^5\,,
\ee
so that we can define chiral projection operators
\be
P_\pm={{\bf 1}\pm\gamma^5\over 2}\,,\label{chiralproj}
\ee
which project the Dirac spinor into two different chiral spinors of the dimension $2^{n-1}$ for each: $\psi=\psi_++\psi_-$.
In the chiral basis where this decomposition is diagonal, that is,
\be
\psi=\left(\begin{array}{c}\psi_+\\\psi_-\end{array}\right)\,,
\ee
we define $2^{n-1}\times 2^{n-1}$ matrices $\sigma_\pm^\mu$ by
\be
P_+\left(-\gamma^0\gamma^\mu\right)P_+=\left(\begin{array}{c|c}\sigma_+^\mu & {\bf 0}\\\hline {\bf 0}& {\bf 0}\end{array}\right)\,,\quad
P_-\left(-\gamma^0\gamma^\mu\right)P_-=\left(\begin{array}{c|c}{\bf 0} & {\bf 0}\\\hline {\bf 0}& \sigma_-^\mu \end{array}\right)\,,\label{sigmadef}
\ee
and the Dirac action in terms of its chiral components $\psi_\pm$ becomes
\be
{\cal L}=i\psi_+^\dagger \sigma_+^\mu  \left(\partial_\mu-ieA_\mu\right)\psi_++i\psi_-^\dagger \sigma_-^\mu  \left(\partial_\mu-ieA_\mu\right)\psi_-\,,
\ee
so that one can nicely separate the two chiral components in the action. In the following, we take only $\psi_+$ chiral spinor and omit $+$ subscripts in our notation. Then, our action for the chiral spinor reads simply as
\be
{\cal L}=i\psi^\dagger \sigma^\mu  \left(\partial_\mu-ieA_\mu\right)\psi\,.\label{chiralaction}
\ee
Note that $\sigma^0={\bf 1}_{2^{n-1}\times 2^{n-1}}$, and $\sigma^\mu$ are hermitian. 
The $\sigma^i$ for spatial indices $i=1,\ldots, 2n-1$ satisfy the anti-commutation relations
\be
\{\sigma^i,\sigma^j\}=2\delta^{ij}\,,\label{sigmaij}
\ee
which can be derived from the anti-commutation relations of the $\gamma$ matrices (\ref{anti}).
This will be helpful in the subsequent discussion on the quantization of the chiral spinor. For later convenience, let us define one more object $\bar \sigma^\mu_\pm$ by
\be
 P_+\left(-\gamma^\mu\gamma^0\right)P_+=\left(\begin{array}{c|c}\bar \sigma_+^\mu & {\bf 0}\\\hline {\bf 0}& {\bf 0}\end{array}\right)\,,\quad
P_-\left(-\gamma^\mu\gamma^0\right)P_-=\left(\begin{array}{c|c}{\bf 0} & {\bf 0}\\\hline {\bf 0}& \bar \sigma_-^\mu \end{array}\right)\,,\label{sigmabardef}
\ee
which satisfy (omitting $+$ subscript again)
\be
\sigma^\mu\bar\sigma^\nu+\sigma^\nu\bar\sigma^\mu=-2\eta^{\mu\nu}\,,\quad \bar\sigma^0=\sigma^0={\bf 1}\,,\quad \bar\sigma^i=-\sigma^i\,.
\ee
A usefulness of $\bar\sigma^\mu$ is from the equation
\be
(p\cdot\sigma)(p\cdot\bar\sigma)=-p^2\,,
\ee
where $(p\cdot\sigma)=p_\mu \sigma^\mu$ for any Lorentz vector $p$, so that the inverse of $(p\cdot\sigma)$ is given by
\be
{1\over (p\cdot\sigma)}=-{(p\cdot\bar\sigma)\over p^2}\,.
\ee

Let us quantize our chiral spinor field. 
The equal time commutation relation from the action (\ref{chiralaction}) is
\be
\{\psi^\dagger_\beta(\vec x),\psi_\alpha(\vec y)\}=\delta^{(d)}(\vec x-\vec y)\delta_{\alpha\beta}\,,\label{basiccom}
\ee
where the Greek letters run over spinor indices, and the operator equation of motion in the free theory is
\be
\sigma^\mu \partial_\mu\psi =0\,.
\ee
The classical spinors satisfying the same equation of motion in the momentum space $p^\mu=(\omega, \vec p)$ divide into two categories depending on the sign of the energy $p^0=\omega=\pm|\vec p|$: 

1) Positive particle states ($\omega=+|\vec p|$)
\be
{\vec\sigma\cdot\vec p\over |\vec p|} u^s(\vec p)=u^s(\vec p)\,,\quad s=1,\cdots, 2^{n-2}\,,
\ee
where $s$ denotes $2^{n-2}$ degenerate spin states. 

2) Negative anti-particle states ($\omega=-|\vec p|$)
\be
{\vec\sigma\cdot\vec p\over |\vec p|} v^s(\vec p)=-v^s(\vec p)\,,\quad s=1,\cdots, 2^{n-2}\,.
\ee
Because $(\vec\sigma\cdot\vec p)$ is hermitian with $(\vec\sigma\cdot\vec p)^2=|\vec p|^2$ (see (\ref{sigmaij})), and ${\rm Tr}(\vec \sigma)=0$ (from the definition (\ref{sigmadef})),
the classical spinors $u^s(\vec p), v^s(\vec p)$ which are eigenvectors of $(\vec\sigma\cdot\vec p)$ span the whole $2^{n-1}$ dimensional chiral spinor space.
It is also convenient to introduce projection operators to the positive and negative energy states by (not to be confused with chiral projection operators (\ref{chiralproj}))
\be
{\cal P}_\pm={1\over 2}\left({\bf 1}\pm {\vec\sigma\cdot\vec p\over |\vec p|}\right)\,.\label{energyproj}
\ee
We choose to normalize the spinors $u^s(\vec p),v^s(\vec p)$ such that 
\be
\sum_s u^{s\dagger}_\beta(\vec p) u^s_\alpha(\vec p) = 2 |\vec p|\left({\cal P}_+\right)_{\alpha\beta}\,,\quad
\sum_s v^{s\dagger}_\beta(\vec p) v^s_\alpha(\vec p) = 2 |\vec p|\left({\cal P}_-\right)_{\alpha\beta}\,.\label{normalization}
\ee
With these, the quantized chiral spinor operator is realized as
\be
\psi(\vec x, t)=\int {d^d \vec p\over (2\pi)^d \sqrt{2|\vec p|}}\sum_s\left(a^s_{\vec p}\,e^{-i|\vec p|t+i\vec p\cdot\vec x} \,u^s(\vec p)+b^{s\dagger}_{-\vec p} \,e^{i|\vec p|t-i\vec p\cdot\vec x}\,v^s(-\vec p)\right)\,,\label{quantumpsi}
\ee
with annihilation operators of particles and anti-particles, $(a^s_{\vec p}$, $b^s_{\vec p})$, respectively, which satisfy the usual anti-commutation relations
\be
\{a^s_{\vec p}\,,\, a^{s'\dagger}_{\vec p'}\}=(2\pi)^d \delta^{(d)}(\vec p-\vec p')\delta^{s s'}\,,\quad
\{b^s_{\vec p}\,,\, b^{s'\dagger}_{\vec p'}\}=(2\pi)^d \delta^{(d)}(\vec p-\vec p')\delta^{s s'}\,.\label{creationcom}
\ee
It is straightforward to check (\ref{basiccom}) using (\ref{normalization}). The Hamiltonian is computed as
\be
{\cal H}=-i\int d^d\vec x\,\psi^\dagger(\vec x)(\vec\sigma\cdot\vec\partial )\psi(\vec x)=
\int {d^d\vec p\over (2\pi)^d}\sum_s |\vec p|\left(a^{s\dagger}_{\vec p} a^s_{\vec p}+b^{s\dagger}_{\vec p} b^s_{\vec p}\right)\,,\label{H}
\ee
up to normal ordering as expected.

We will be interested in the expectation values of operators and correlation functions at a finite temperature $T$ and a chemical potential $\mu$.
The thermal ensemble is defined as usual
\be
\langle {\cal O}\rangle \equiv {{\rm Tr}\left(e^{-\beta\left({\cal H}-\mu {\cal N}\right)} {\cal O}\right)\over {\rm Tr}\left(e^{-\beta\left({\cal H}-\mu{\cal N}\right)}\right)}\,,\label{thermalop}
\ee
 where 
 \be
 {\cal N}=\int {d^d\vec p\over (2\pi)^d}\sum_s\left(a^{s\dagger}_{\vec p} a^s_{\vec p}-b^{s\dagger}_{\vec p} b^s_{\vec p}\right)\,.\label{N}
 \ee
 With (\ref{creationcom}), (\ref{H}), and (\ref{N}), one can show that
 \bear
 \langle a^{s\dagger}_{\vec p} a^{s'}_{\vec p'}\rangle &=& \delta^{s s'}(2\pi)^d \delta^{(d)}(\vec p-\vec p'){1\over e^{\beta\left(|\vec p|-\mu\right)}+1}\,,\\\label{aadagger}
  \langle b^{s\dagger}_{\vec p} b^{s'}_{\vec p'}\rangle& =& \delta^{s s'}(2\pi)^d \delta^{(d)}(\vec p-\vec p'){1\over e^{\beta\left(|\vec p|+\mu\right)}+1}\,,\label{bbdagger}
 \eear
 which, in conjunction with (\ref{quantumpsi}), allow us to compute any kind of two point correlation functions of $\psi$ and $\psi^\dagger$.

\section{Diagrammatic computation of real-time retarded functions}

\begin{figure}[t]
	\centering
	\includegraphics[width=8cm]{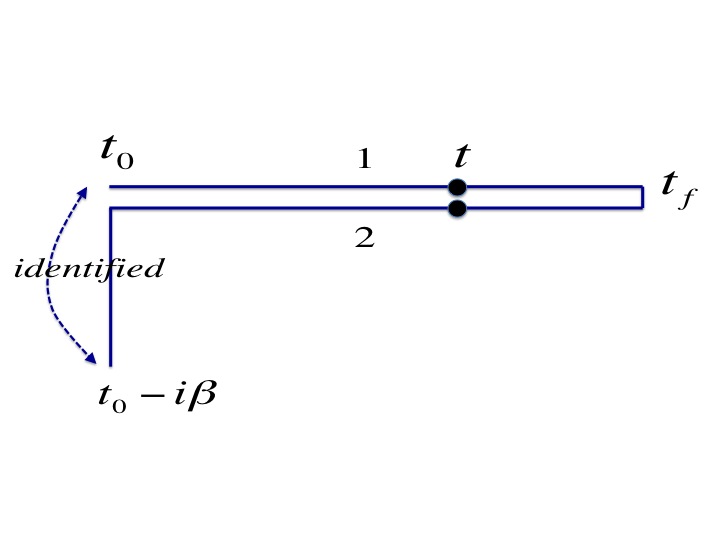}
		\caption{The Schwinger-Keldysh contour appropriate for computing real-time retarded response functions at finite temperature.\label{fig1}}
\end{figure}

What we are interested in is the current induced by the external $U(1)$ gauge field coupled to the number current $J^\mu=\psi^\dagger\sigma^\mu\psi$.
The action including the interaction with the external gauge field is
\be
{\cal L}=i\psi^\dagger\sigma^\mu\partial_\mu\psi +e\left(\psi^\dagger\sigma^\mu\psi\right) A_\mu\,,\label{action1}
\ee
and we are going to do a perturbation expansion in $eA_\mu$.
Since we are going to compute the thermal expectation value of an operator, $J^\mu$, in the presence of $A_\mu$, one naturally introduces the Schwinger-Keldysh 
contour in the complex time plane as shown in Figure \ref{fig1} in the path-integral formalism. 
We will discuss the translation of this path integral formalism to our operator formalism in the previous section.  In simple terms,
the upper line (the real-time line labeled as 1) represents the unitary time evolution of the ket state 
\be
|t\rangle=U(t,t_0)|t_0\rangle\,,\quad U(t,t_0)={\cal P}e^{-i\int^t_{t_0} {\cal H}(t') dt'}\,,
\ee
whereas the lower line labeled as 2 describes the time evolution of the bra state, the conjugate state of the ket state,
\be
\langle t|=\langle t_0|U^\dagger(t,t_0)=\langle t_0|U(t_0,t)\,,
\ee 
so that the resulting path integral with an operator, say $J^\mu$, inserted at a time $t$ naturally calculates the expectation value 
\be
J^\mu(t)=\langle t|J^\mu |t\rangle=\langle t_0| U(t_0,t) J^\mu U(t,t_0)|t_0\rangle\,.
\ee
Note that the evolution matrix $U(t_0,t)$ for the bra state is a time-reversed one, and this is why the action for the contour line 2 in the Schwinger-Keldysh path integral
is the negative of the ordinary action (\ref{action1}):
\be
{\cal L}_2=-i\psi^\dagger_2\sigma^\mu\partial_\mu\psi_2 -e\left(\psi^\dagger_2\sigma^\mu\psi_2\right) A_\m\,,
\ee
where we put a subscript 2 in the dynamical fields for clarity. 
Note also that the path integral on the time interval greater than $t$ (the part of the contour on the right of the operator $J^\mu$ inserted) cancels between the lines 1 and 2,
if our boundary condition at the final time $t_f$ is such that $\psi_1(t_f)=\psi_2(t_f)$, since the two evolution operators $U(t_f,t)$ and $U(t,t_f)$ generated by the lines 1 and 2 respectively are precisely inverse to each other. 
This automatically guarantees the causal response of the current expectation value $J^\mu(t)$ to the perturbation $A_\mu$, since the $A_\mu(t')$ for $t'>t$ which appears 
on the right hand side of the contour from the $J^\mu$ insertion at $t$ would never affect the resulting path integral for $J^\mu(t)$.
In other words, $J^\mu(t)$ computed in the Schwinger-Keldysh path integral in a perturbation expansion in $eA_\mu$
gives us a series of retarded causal $n$-point real-time response functions of the currents by construction. In the notation that will be introduced soon, they are $
G_{ra\cdots a}$ correlation functions of the current.
We stress that this is crucially based on the continuous boundary condition at the final time $t_f$. The far left part of the contour in Figure \ref{fig1} is responsible for the thermal ensemble by circling around the imaginary time of a period $\beta=T^{-1}$ as usual. The causality discussed above and the naturalness of having the two contours 1 and 2 for bra and ket states for any expectation values of operators do not depend on what ensemble we consider, and are more generic. In this sense, introducing the Schwinger-Keldysh contour
with a continuous boundary condition at the final time $t_f$ is an inevitable step in computing retarded response functions. 

The free theory Schwinger-Keldysh path integral is entirely Gaussian, so that the Wick theorem holds true for free theory correlation functions, which allows one to apply the Feynman diagram techniques in any perturbation theory from the free limit in computing retarded response functions in thermal equilibrium: this is the essence of the formalism which may look highly non-trivial in the language of operator formalism since we are dealing with thermal ensemble expectation values.

The path integral measure from the two contour lines 1 and 2 is
\be
\exp\left[i\int_{t_0}^{t_f} \left({\cal L}_1+{\cal L}_2\right)\right]=\exp\left[\int_{t_0}^{t_f}\left(-\psi_1^\dagger\sigma^\mu\partial_\mu\psi_1+i e \psi_1^\dagger\sigma^\mu\psi_1 A_\mu+\psi_2^\dagger\sigma^\mu\partial_\mu\psi_2-ie \psi_2^\dagger\sigma^\mu\psi_2 A_\mu\right)\right]\,,\label{SKaction}
\ee
where we skip the the Euclidean path integral arising from the far left part of the contour generating the thermal ensemble. We can assume that the gauge field vanishes 
at a sufficiently past time $t_0\to-\infty$, so that this Euclidean path integral part does not contain any external gauge field $A_\mu$: the thermal ensemble is the one in the free theory that we discuss in the previous section.
The current expectation value of our interest is simply the path integral
\be
J^\mu(t)=\langle \psi_1^\dagger\sigma^\mu\psi_1(t)\rangle_{\rm SK}=\langle \psi_2^\dagger\sigma^\mu\psi_2(t)\rangle_{\rm SK}\,,
\ee
where $\langle\cdots\rangle_{\rm SK}$ is the path integral with the Schwinger-Keldysh contour (not to be confused with the operator expectation value in (\ref{thermalop})).
Note that it does not matter in the above whether we put $\psi_1^\dagger\sigma^\mu\psi_1$ or $ \psi_2^\dagger\sigma^\mu\psi_2$, since the part of the contour with $t'>t$ cancels by itself. To do a perturbation theory in $eA_\mu$ it is convenient to work in the ``ra'' combinations defined by
\be
\psi_r\equiv {1\over 2}\left(\psi_1+\psi_2\right)\,,\quad \psi_a\equiv \psi_1-\psi_2\,,
\ee
in terms of which the action in (\ref{SKaction}) becomes
\be
\exp\left[i\int_{t_0}^{t_f} \left({\cal L}_1+{\cal L}_2\right)\right]=\exp\left[\int_{t_0}^{t_f} \left(-\psi_a^\dagger\sigma^\mu\partial_\mu\psi_r -\psi_r^\dagger\sigma^\mu\partial_\mu\psi_a
+ie \left(\psi_a^\dagger\sigma^\mu\psi_r+\psi_r^\dagger\sigma^\mu\psi_a\right)A_\mu\right)\right]\,,\label{raaction}
\ee
and the current we insert for the expectation value can be chosen as
\be
J^\mu_r={1\over 2}\left(\psi_1^\dagger\sigma^\mu \psi_1+\psi_2^\dagger\sigma^\mu \psi_2\right)=\psi_r^\dagger\sigma^\mu\psi_r+{1\over 4}\psi_a^\dagger\sigma^\mu\psi_a\,.\label{jmurtype}
\ee
One can find that the second piece does not contribute anything in the expectation value, so can be ignored. The usefulness of the above ``ra''-basis is
due to the boundary condition at $t_f$: $\psi_a(t_f)=0$. From the structure of the free theory action in the ra-basis, this ensures that  any free theory correlation function with an ``a''-type operators
appearing at the latest time always vanishes: this holds true for two point functions trivially, and the Wick theorem generalizes it to arbitrary correlation functions.
This property is nothing but what ensures the causal response as discussed before in a different language, since the external perturbation such as $A_\mu$ couples precisely to an ``a''-type operator. On the other hand, the physical expectation value is computed by the ``r''-type operator as shown in (\ref{jmurtype}).
This means that the causal $n$-point response functions are the correlation functions of the type $G_{ra\cdots a}$ where the physical observable corresponds to the first ``r'' and the operators coupling to the external perturbations belong to the other ``a''-types.

It is straightforward to write down the Feynman rules for the perturbation theory from the action (\ref{raaction}) in the ra-basis. The basic building block two-point functions are
defined as follows,
\bear
G_{ra}(x,y)&=&\langle \psi_r(x)\psi^\dagger_a(y)\rangle_{\rm SK}=\int {d^{2n} p\over (2\pi)^{2n}} e^{ip\cdot(x-y)}G_{ra}(p)\,,\\
G_{ar}(x,y)&=&\langle \psi_a(x)\psi^\dagger_r(y)\rangle_{\rm SK}=\int {d^{2n} p\over (2\pi)^{2n}} e^{ip\cdot(x-y)}G_{ar}(p)\,,\\
G_{rr}(x,y)&=&\langle \psi_r(x)\psi^\dagger_r(y)\rangle_{\rm SK}=\int {d^{2n} p\over (2\pi)^{2n}} e^{ip\cdot(x-y)}G_{rr}(p)\,,
\eear
where the both sides should be understood as $2^{n-1}\times 2^{n-1}$ matrices of spinor indices we omit here, and $x,y$ are $d+1=2n$ dimensional space-time coordinates.
Note that $G_{aa}$ is absent. To compute above two point functions explicitly, we translate them into the operator formalism so that we can use the results in the previous section.
Considering operator time ordering carefully, one can indeed show that
\bear
G_{ra}(x,y)&=&\theta(x^0-y^0)\langle \{\psi(x),\psi^\dagger(y)\}\rangle\,,\\
G_{ar}(x,y)&=&-\theta(y^0-x^0)\langle \{\psi(x),\psi^\dagger(y)\}\rangle\,,\\
G_{rr}(x,y)&=& {1\over 2}\langle[\psi(x),\psi^\dagger(y)]\rangle\,,
\eear
where $\langle\cdots\rangle$ is the operator thermal ensemble average introduced in (\ref{thermalop}).
For example, the equation for $G_{ra}$ is derived as follows,
\bear
2\,G_{ra}(x,y)&=& \langle \psi_1(x)\psi^\dagger_1(y)\rangle_{\rm SK}+\langle \psi_2(x)\psi^\dagger_1(y)\rangle_{\rm SK}-\langle \psi_1(x)\psi_2^\dagger(y)\rangle_{\rm SK}-\langle \psi_2(x)\psi_2^\dagger (y)\rangle_{\rm SK}\nonumber\\
&=& \langle {\cal T}\psi(x)\psi^\dagger(y)\rangle +\langle \psi(x)\psi^\dagger(y)\rangle+\langle \psi^\dagger(y)\psi(x)\rangle-\langle\bar{\cal T}\psi(x)\psi^\dagger(y)\rangle\nonumber\\
&=& 2\,\theta(x^0-y^0)\langle \{\psi(x),\psi^\dagger(y)\}\rangle\,,
\eear
where $\cal T$ and $\bar {\cal T}$ are time ordering and anti-time ordering respectively. We see that the $G_{ra}$ is the retarded two point function and $G_{ar}$ is the advanced one. 
The $G_{rr}$ encodes thermal fluctuations. Using the quantum expansion (\ref{quantumpsi}) and the explicit thermal expectation values (\ref{aadagger}) and (\ref{bbdagger}), it is straightforward to compute the above two point functions  after some amount of algebra to obtain
\bear
G_{ra}(p)&=&{i\over p^0-|\vec p|+i\epsilon}{\cal P}_++{i\over p^0+|\vec p|+i\epsilon}{\cal P}_-=i{p^0{\bf 1}+\vec\sigma\cdot\vec p\over (p^0+i\epsilon)^2-|\vec p|^2}={-i (p\cdot\bar\sigma)\over (p^0+i\epsilon)^2-|\vec p|^2}\,, \nonumber\\
G_{ar}(p)&=&{i\over p^0-|\vec p|-i\epsilon}{\cal P}_++{i\over p^0+|\vec p|-i\epsilon}{\cal P}_-={-i (p\cdot\bar\sigma)\over (p^0-i\epsilon)^2-|\vec p|^2}\,, \nonumber\\
G_{rr}(p)&=& -{\pi\over|\vec p|}\left(\delta\left(p^0-|\vec p|\right)-\delta\left(p^0+|\vec p|\right)\right)\left({1\over 2}-n_+(p^0)\right)\left(\bar\sigma\cdot p\right)\,,
\eear
where the projection operators ${\cal P}_\pm$ are defined as before in (\ref{energyproj}),
\be
{\cal P}_\pm={1\over 2}\left({\bf 1}\pm {\vec\sigma\cdot\vec p\over |\vec p|}\right)\,,
\ee
and 
\be
n_+(p^0)={1\over 1+e^{\beta(p^0-\mu)}}\,,
\ee
is the thermal distribution with chemical potential $\mu$. Note that $G_{ra}$ and $G_{ar}$ do not depend on temperature in the free theory, since $\{\psi(x),\psi^\dagger(y)\}$ is proportional to the identity operator for any $(x,y)$. 

\begin{figure}[t]
	\centering
	\includegraphics[width=8cm]{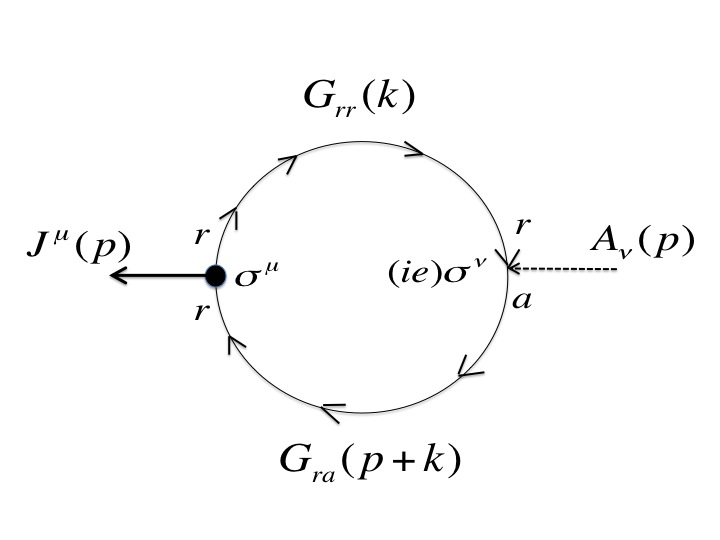}\includegraphics[width=8cm]{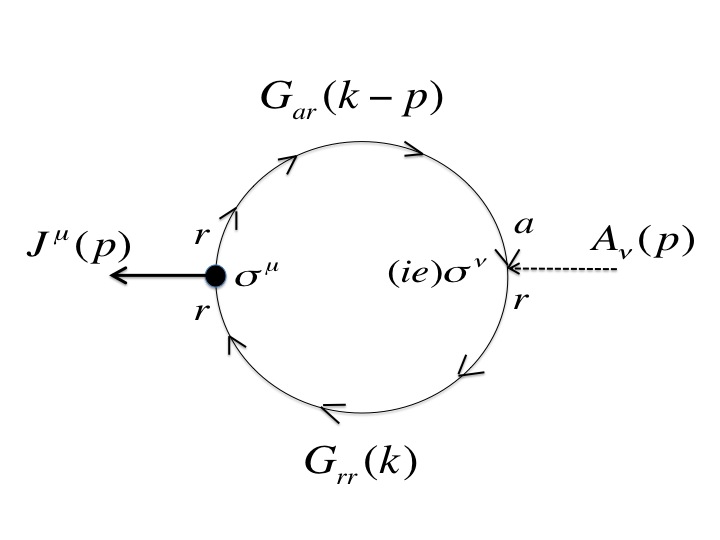}
		\caption{The diagrams responsible for the retarded response of the current $J^\mu$ to one external gauge potential.\label{fig3}}
\end{figure}

In the Feynman diagrams in momentum space, each fermion line is drawn with an arrow whose direction is from $\psi^\dagger$ to $\psi$.
For simplicity, we choose the same arrow to also mean the momentum direction carried by the fermion line. 
In writing down the expression corresponding to a diagram, one writes the terms from right to left when following the arrow direction.
Each fermion loop accompanies an extra $(-1)$ sign after the spinor trace. Each loop integral measure is
\be
\int {d^{2n} k\over (2\pi)^n}\,.
\ee
From the form of the action (\ref{raaction}), 
each external gauge field with momentum $p$, $A_\mu(p)$, gives a vertex insertion $(ie)\,\sigma^\mu$, either ``ra'' or ``ar'' type.
What we are going to compute is the expectation value of the current (in momentum space)
\be
J^\mu_r=\psi_r^\dagger\sigma^\mu\psi_r+{1\over 4}\psi_a^\dagger\sigma^\mu\psi_a\,,
\ee
where one can easily convince oneself that there is no possible diagram involving the second term, so we can consider only the first term.
As an example, let's consider the causal response of $J^\mu$ which are linear in the external gauge potential (and hence we should consider diagrams with two currents inserted).
As shown in Figure \ref{fig3}, there are two diagrams possible. The first diagram involves $G_{ra}$ and $G_{rr}$, whereas the second diagram contains $G_{ar}$ and $G_{rr}$.
We choose our loop momentum such that the momentum appearing in the $G_{rr}$ line is always $k$. 
Then, the resulting expression for $J^\mu(p)$ is
\be
J^\mu_{(1)}(p)=(-1)ie A_\nu(p)\int {d^{2n}k\over (2\pi)^{2n}}{\rm tr}\left[\sigma^\mu G_{ra}(p+k)\sigma^\nu G_{rr}(k)+\sigma^\mu G_{rr}(k)\sigma^\nu G_{ar}(k-p)\right]\,.
\ee

\begin{figure}[t]
	\centering
	\includegraphics[width=10cm]{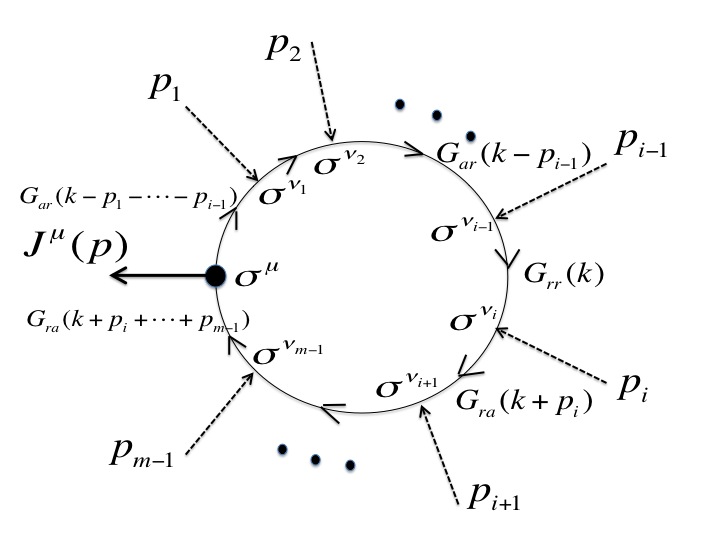}
		\caption{One of the diagrams responsible for the retarded response of the current $J^\mu$ to $(m-1)$ number of external gauge potentials. There are $m$ number of such diagrams with different positions of $G_{rr}$ propagator.\label{fig4}}
\end{figure}
We now discuss the general structure of the diagrams with $(m-1)$ number of external gauge potentials ($m>1$). 
There are $m$ number of possible diagrams, which is organized as follows. Each diagram is a 1-loop diagram with $m$ number of currents inserted, and one of them is $J^\mu(p)$.
We call the external momentum of the $i$'th attached gauge field, $A_{\nu_i}$, labelled from $J^\mu$ along the arrow direction, $p_i$, $i=1,\ldots (m-1)$, so that there are $(m-1)$ vertex insertions $ie A_{\nu_i}(p_i)\sigma^{\nu_i}$, $i=1,\ldots (m-1)$. There is an overall $\delta$ function dictating the momentum conservation, $(2\pi)^{2n}\delta^{(2n)}\left(p-p_1-\cdots p_{m-1}\right)$, as usual.
Among the $m$ number of fermion lines, one can choose one line to be $G_{rr}$ with the loop momentum $k$. Then to have a non-vanishing diagram,  all the fermion lines along the arrow direction between $J^\mu$ and the chosen line should be $G_{ar}$, and all the fermion lines from the chosen line to the $J^\mu$ insertion must be $G_{ra}$: the diagram is uniquely determined by the position of the $G_{rr}$ line in the loop. There are precisely $m$ number of ways to have different diagrams. Figure \ref{fig4} shows the diagram where the $G_{rr}$ line is located between $(i-1)$'th and $i$'th gauge potential insertions, $i=1,\ldots,m$ ($0$'th and $m$'th insertion are by definition $J^\mu(p)$ itself).
This diagram gives
\bear
&&(-1)(ie)^{m-1}\int{d^{2n}p_1\over (2\pi)^{2n}}\cdots\int{d^{2n}p_{m-1}\over (2\pi)^{2n}} \,\,(2\pi)^{2n}\delta\left(p-p_1-\cdots-p_{m-1}\right)A_{\nu_1}(p_1)\cdots A_{\nu_{m-1}}(p_{m-1})\nonumber\\ &&\int {d^{2n}k\over (2\pi)^{2n}}\,\,{\rm tr}\bigg[\sigma^\mu G_{ra}(k+p_i+\cdots+p_{m-1})\sigma^{\nu_{m-1}}\cdots G_{ra}(k+p_i)\sigma^{\nu_i}G_{rr}(k)\nonumber\\ &&\times \sigma^{\nu_{i-1}}G_{ar}(k-p_{i-1})\cdots\sigma^{\nu_1}G_{ar}(k-p_1-\cdots p_{i-1})\bigg] \,,\label{ith}
\eear
and we have to sum over $i=1,\ldots,m$ to find the final $J^\mu_{(m-1)}(p)$ in $(m-1)$'th order of the gauge potential. When $i=1$ ($m$), the $G_{ar}$ ($G_{ra}$) are absent in the above formula.

\section{Chiral magnetic effect in $d=2n$ dimensions}

As discussed in the introduction, the CME in $2n$ dimensions appears in the $(n-1)$'th order of the external gauge field, so that we have to compute $m=n$ number of
diagrams whose contributions are given by (\ref{ith}) with $m=n$ and $i=1,\ldots,n$. 
In general, the result is highly non-analytic near the zero momenta $p_i\to 0$ region, so that the result in the zero momentum limit will in general depend on how
one approaches the zero momentum. Guided by previous observations in literature, we expect that the correct CMW coefficient is obtained when we first let the frequencies 
be zero, $p^0_i\to 0$, before taking the zero spatial momentum limit, $\vec p_i\to 0$. In this section, we therefore compute (\ref{ith}) after taking $p^0_i\to 0$ limit, and show that
one indeed recovers the right magnitude of the CME in $2n$ dimensions in this limit.  The computation of (\ref{ith}) simplifies greatly in this zero frequency limit, $p_i^0\to 0$, which allows us some degree of analytic computations. In this limit, one can also map the problem to the purely Euclidean computation, but we will skip persueing this possibility.


We aim to compute the loop integral in (\ref{ith}) with $m=n$,
\bear
&&\int {d^{2n}k\over (2\pi)^{2n}}\,\,{\rm tr}\bigg[\sigma^\mu G_{ra}(k+p_i+\cdots+p_{n-1})\sigma^{\nu_{n-1}}\cdots G_{ra}(k+p_i)\sigma^{\nu_i}G_{rr}(k)\nonumber\\ &&\times \sigma^{\nu_{i-1}}G_{ar}(k-p_{i-1})\cdots\sigma^{\nu_1}G_{ar}(k-p_1-\cdots p_{i-1})\bigg]\,,\label{ithloop}
\eear
and sum the result over $i=1,\ldots,m$.
The numerator of the integrand is
\bear
&&(-i)^{n-1}{\rm tr}\bigg[\sigma^\mu \left((k+p_i+\cdots p_{n-1})\cdot\bar\sigma\right)\sigma^{\nu_{n-1}}\cdots \left((k+p_i)\cdot\bar\sigma\right)\sigma^{\nu_i}\left(k\cdot\bar\sigma\right)\nonumber\\&&\times \sigma^{\nu_{i-1}}\left((k-p_{i-1})\cdot\bar\sigma\right)\cdots\sigma^{\nu_1} \left((k-p_1-\cdots-p_{i-1})\cdot\bar\sigma\right)\bigg]\,.\label{numerator1}
\eear
We are interested in only the P-odd part of the contribution which involves the $\epsilon$-tensor, and we need to use the following statement that the P-odd part of the trace
\be
{\rm tr}\left[\sigma^{\mu_1}\bar\sigma^{\nu_1}\sigma^{\mu_2}\bar\sigma^{\nu_2}\cdots \sigma^{\mu_n}\bar\sigma^{\nu_n}\right]\,,\label{sigmatrace}
\ee
is given by
\be
(2i)^{n-1}\,\,\epsilon^{\mu_1\nu_1\cdots\mu_n\nu_n}\,,
\ee
where by definition, $\epsilon^{012\cdots (2n-1)}=+1$. To show this, start from the definitions (\ref{sigmadef}) and (\ref{sigmabardef}) to have
\bear
&&{\rm tr}\left[\sigma^{\mu_1}\bar\sigma^{\nu_1}\sigma^{\mu_2}\bar\sigma^{\nu_2}\cdots \sigma^{\mu_n}\bar\sigma^{\nu_n}\right]=
{\rm tr}\left[P_+\left(\gamma^0\gamma^{\mu_1}\gamma^{\nu_1}\gamma^0\right)\cdots \left(\gamma^0\gamma^{\mu_n}\gamma^{\nu_n}\gamma^0\right)\right]\nonumber\\
&&=(-1)^n\,\,{\rm tr}\left[P_-\gamma^{\mu_1}\gamma^{\nu_1}\cdots \gamma^{\mu_n}\gamma^{\nu_n}\right]={(-1)^n\over 2}\,\,{\rm tr}\left[({\bf 1}-\gamma^5)\gamma^{\mu_1}\gamma^{\nu_1}\cdots \gamma^{\mu_n}\gamma^{\nu_n}\right]\,.
\eear
The P-odd part is obtained from the $\gamma^5$ matrix, and using the fact that $\gamma^5=i^{n-1}\,\,\gamma^0\gamma^1\cdots\gamma^{2n-1}$, one has
\be
-{1\over 2}(-1)^n\,\,{\rm tr}\left[\gamma^5\gamma^{\mu_1}\gamma^{\nu_1}\cdots \gamma^{\mu_n}\gamma^{\nu_n}\right]={1\over 2}(i)^{n-1}\epsilon^{\mu_1\nu_1\cdots\mu_n\nu_n}{\rm tr}\,\,{\bf 1}=(2i)^{n-1}\,\,\epsilon^{\mu_1\nu_1\cdots\mu_n\nu_n}\,.
\ee
Using this, the P-odd part of the numerator (\ref{numerator1}) becomes after some algebra
\be
2^{n-1}k_\nu(p_1)_{\mu_1}\cdots(p_{n-1})_{\mu_{n-1}}\,\,\epsilon^{\mu\nu\mu_1\nu_1\cdots\mu_{n-1}\nu_{n-1}}\,,\label{nume1}
\ee
which is the same for all $i=1,\ldots,m$.

What is difficult is the rest part including the denominator of the integrand. It is written as
\bear
&&\left(-{\pi\over|\vec k|}\right)\left(\delta(k^0-|\vec k|)-\delta(k^0+|\vec k|)\right)\left({1\over 2}-n_+(k^0)\right)\nonumber\\
&\times&{1\over \left[(k^0+i\epsilon)^2-|\vec k+\vec p_i+\cdots+\vec p_{n-1}|^2\right]\cdots\left[(k^0+i\epsilon)^2-|\vec k+\vec p_i|^2\right]}\nonumber\\
&\times& {1\over\left[(k^0-i\epsilon)^2-|\vec k-\vec p_{i-1}|^2\right]\cdots\left[(k^0-i\epsilon)^2-|\vec k-\vec p_1-\cdots-\vec p_{i-1}|^2\right]}\,,\label{deno0}
\eear
where we have put all $p^0_i=0$. Using the on-shellness, $(k^0)^2=|\vec k|^2$, given by the delta functions, the above becomes
\bear
&&\left(-{\pi\over|\vec k|}\right)\left(\delta(k^0-|\vec k|)-\delta(k^0+|\vec k|)\right)\left({1\over 2}-n_+(k^0)\right)\label{deno1}\\
&\times&{1\over \left[-2\vec k\cdot(\vec p_i+\cdots+\vec p_{n-1})-|\vec p_i+\cdots+\vec p_{n-1}|^2+i k^0\epsilon\right]\cdots\left[-2\vec k\cdot\vec p_i-|\vec p_i|^2+ik^0\epsilon\right]}\nonumber\\
&\times& {1\over\left[2\vec k\cdot\vec p_{i-1}-|\vec p_{i-1}|^2-ik^0\epsilon\right]\cdots\left[2\vec k\cdot(\vec p_1+\cdots+\vec p_{i-1})-|\vec p_1+\cdots+\vec p_{i-1}|^2-i k^0\epsilon\right]}\,.\nonumber
\eear
There are total $(n-1)$ terms in the denominator, and we would like to combine them using the Feynman parametrization
\be
{1\over (A_1\pm i\epsilon)\cdots(A_{n-1}\pm i\epsilon)}=(n-2)!\int^1_0\,dx_1\cdots dx_{n-1} {\delta(1-x_1-\cdots-x_{n-1})\over (x_1A_1+\cdots+x_{n-1}A_{n-1}\pm i\epsilon)^{n-1}}\,.
\ee
It is worth emphasizing that the Feynman formula is valid only with the crucial presence of $i\epsilon$ in each term with the overall {\it same} sign: if some of $i\epsilon$ term appears with a different sign compared to others, the formula is not valid.
Looking at (\ref{deno1}), we see that the first $(n-i)$ terms have $+ik^0\epsilon$ while the rest $(i-1)$ terms have $-ik^0\epsilon$. To make $i\epsilon$ terms having the same sign,
we consider minus of each of the first $(n-i)$ terms to have
\bear
&&(-1)^{n-i}\left(-{\pi\over|\vec k|}\right)\left(\delta(k^0-|\vec k|)-\delta(k^0+|\vec k|)\right)\left({1\over 2}-n_+(k^0)\right)\label{deno2}\\
&\times&{1\over \left[2\vec k\cdot(\vec p_i+\cdots+\vec p_{n-1})+|\vec p_i+\cdots+\vec p_{n-1}|^2-i k^0\epsilon\right]\cdots\left[2\vec k\cdot\vec p_i+|\vec p_i|^2-ik^0\epsilon\right]}\nonumber\\
&\times& {1\over\left[2\vec k\cdot\vec p_{i-1}-|\vec p_{i-1}|^2-ik^0\epsilon\right]\cdots\left[2\vec k\cdot(\vec p_1+\cdots+\vec p_{i-1})-|\vec p_1+\cdots+\vec p_{i-1}|^2-i k^0\epsilon\right]}\,,\nonumber
\eear
which now has the overall same sign for $i\epsilon$'s in each term, so that one can safely use the Feynman formula.
The result is
\bear
&&(-1)^{n-i}\left(-{\pi\over|\vec k|}\right)\left(\delta(k^0-|\vec k|)-\delta(k^0+|\vec k|)\right)\left({1\over 2}-n_+(k^0)\right)\nonumber\\
&\times& (n-2)!\int^1_0 \,dx_1\cdots dx_{n-1}\,{\delta(1-x_1-\cdots-x_{n-1})\over \left[\vec k\cdot\vec Q_i+\Delta_i-ik^0\epsilon\right]^{n-1}}\,\,,\label{deno3}
\eear
where
\bear
\vec Q_i&=&2\left(x_{n-1}(\vec p_i+\cdots+\vec p_{n-1})+\cdots+x_i\vec p_i+x_{i-1}\vec p_{i-1}+\cdots +x_1(\vec p_1+\cdots+\vec p_{i-1})\right)\,,\nonumber\\
\Delta_i&=& x_{n-1}|\vec p_i+\cdots+\vec p_{n-1}|^2+\cdots+x_i|\vec p_i|^2-x_{i-1}|\vec p_{i-1}|^2-\cdots-x_1|\vec p_1+\cdots+\vec p_{i-1}|^2\,.\nonumber\\
\eear
As  examples, for $n=3$ we have
\bear
\vec Q_1&=&2\left(x_2(\vec p_1+\vec p_2)+x_1\vec p_1\right)\,,\quad \Delta_1=x_2 |\vec p_1+\vec p_2|^2+x_1|\vec p_1|^2\,,\nonumber\\
\vec Q_2&=&2\left(x_2\vec p_2+x_1\vec p_1\right)\,,\quad \Delta_2=x_2 |\vec p_2|^2-x_1|\vec p_1|^2\,,\nonumber\\
\vec Q_3&=&2\left(x_2\vec p_2+x_1(\vec p_1+\vec p_2)\right)\,,\quad \Delta_3=-x_2 |\vec p_2|^2-x_1|\vec p_1+\vec p_2|^2\,,
\eear
and for $n=4$ we have
\bear 
\vec Q_1&=&2\left(x_3(\vec p_1+\vec p_2+\vec p_3)+x_2(\vec p_1+\vec p_2)+x_1\vec p_1\right)\,,\nonumber\\
\Delta_1&=&x_3|\vec p_1+\vec p_2+\vec p_3|^2+x_2|\vec p_1+\vec p_2|^2+x_1|\vec p_1|^2\,,\nonumber\\
\vec Q_2&=&2\left(x_3(\vec p_2+\vec p_3)+x_2\vec p_2+x_1\vec p_1\right)\,,\nonumber\\
\Delta_2&=&x_3|\vec p_2+\vec p_3|^2+x_2|\vec p_2|^2-x_1|\vec p_1|^2\,,\nonumber\\
\vec Q_3&=&2\left(x_3\vec p_3+x_2\vec p_2+x_1(\vec p_1+\vec p_2)\right)\,,\nonumber\\
\Delta_3&=&x_3|\vec p_3|^2-x_2|\vec p_2|^2-x_1|\vec p_1+\vec p_2|^2\,,\nonumber\\
\vec Q_4&=&2\left(x_3\vec p_3+x_2(\vec p_2+\vec p_3)+x_1(\vec p_1+\vec p_2+\vec p_3)\right)\,,\nonumber\\
\Delta_4&=&-x_3|\vec p_3|^2-x_2|\vec p_2+\vec p_3|^2-x_1|\vec p_1+\vec p_2+\vec p_3|^2\,.
\eear

Combining (\ref{nume1}) and (\ref{deno3}), our loop integral (\ref{ithloop}) becomes
\bear
&&\pi(-1)^{n}2^{n-1}(n-2)!\,\,(p_1)_{\mu_1}\cdots (p_{n-1})_{\mu_{n-1}}\epsilon^{\mu 0\mu_1\nu_1\cdots\mu_{n-1}\nu_{n-1}}\int^1_0 \prod_{j=1}^{n-1}dx_j\,\,\delta\left(1-\sum_{j=1}^{n-1}x_j\right)\nonumber\\
&\times& \int{d^{2n} k\over (2\pi)^{2n}}\left(\delta(k^0-|\vec k|)+\delta(k^0+|\vec k|)\right)\left({1\over 2}-n_+(k^0)\right){(-1)^i\over  \left[\vec k\cdot\vec Q_i+\Delta_i-ik^0\epsilon\right]^{n-1}}\,,\label{loop2}
\eear
where we have put $\nu=0$ since one can easily check that this is the only non-vanishing possibility for $\nu$ due to anti-symmetric nature of the $\epsilon$-tensor.
At the end of the computation, we have to sum over $i=1,\ldots, n$.

We can now do the loop integration over $k$ as follows: since $\vec Q_i$ is a fixed vector for $\vec k$ integration whose measure is isotropic, one can conveniently choose the direction of
$\vec Q_i$ as $\hat x^{2n-1}$ in the $(2n-1)$ dimensional vector space of $\vec k$. We call the angle between $\vec Q_i$ and $\vec k$ be $\theta$, so that
\be
\vec Q_i\cdot\vec k=|\vec Q_i||\vec k|\cos\theta\,.
\ee
Then the metric in the $\vec k$ space is written as
\be
ds^2=d|\vec k|^2+|\vec k|^2 d\theta^2+|\vec k|^2\sin^2\theta d\Omega_{2n-3}^2\,,
\ee
where $d\Omega^2_{2n-3}$ is the metric on the unit $S^{2n-3}$ sphere. Note that our integrand in the above depends only on $\theta$, so that one can integrate over the $S^{2n-3}$ trivially. Therefore, the measure of the $k$ integration becomes
\bear
\int{d^{2n}k\over (2\pi)^{2n}}&=&{1\over (2\pi)^{2n}}{\rm Vol}\left(S^{2n-3}\right)\int dk^0\int d|\vec k|\,\,|\vec k|^{2n-2}\int^\pi_0 d\theta \,\,\sin^{2n-3}\theta \nonumber\\
&=&{1\over 2^{2n-1}\pi^{n+1}(n-2)!}\int dk^0\int d|\vec k|\,\,|\vec k|^{2n-2}\int^\pi_0 d\theta \,\,\sin^{2n-3}\theta \,.
\eear
Then, the loop integral (\ref{loop2}) reduces to
\bear
&&{(-1)^n\over (2\pi)^n}\,\,(p_1)_{\mu_1}\cdots (p_{n-1})_{\mu_{n-1}}\epsilon^{\mu 0\mu_1\nu_1\cdots\mu_{n-1}\nu_{n-1}}\int^1_0 \prod_{j=1}^{n-1}dx_j\,\,\delta\left(1-\sum_{j=1}^{n-1}x_j\right)\nonumber\\&\times&\int dk^0\int d|\vec k|\,\,|\vec k|^{2n-2}\left(\delta(k^0-|\vec k|)+\delta(k^0+|\vec k|)\right)\left({1\over 2}-n_+(k^0)\right)\nonumber\\
&\times&(-1)^i \int^\pi_0 d\theta\,\,{\sin^{2n-3}\theta\over  \left[|\vec k||\vec Q_i|\cos\theta+\Delta_i-ik^0\epsilon\right]^{n-1}}\,.\label{loop3}
\eear
Since $\Delta_i$ is ${\cal O}\left(|\vec p|^2\right)$ and $|\vec Q_i|$ is ${\cal O}\left(|\vec p|\right)$, we perform a derivative expansion for small $\vec p_i$ limit by expanding the above
integrand in powers of $\Delta_i/|\vec Q_i|$, and try to obtain the first non-zero result after summing over $i=1,\ldots,m$. We will argue that the first non-zero result arises
in the $(n-1)$'th order of the expansion in $\Delta_i/|\vec Q_i|$, based on the following conjecture,
\be
({\rm Conjecture}): \,\, \int^1_0 \prod_{j=1}^{n-1}dx_j\,\,\delta\left(1-\sum_{j=1}^{n-1}x_j\right)\sum_{i=1}^n {(-1)^i\over |\vec Q_i|^{n-1}}\left({\Delta_i\over |\vec Q_i|}\right)^s =0\,,\quad s=0,1,\ldots, (n-2)\,.\label{conj1}
\ee
We couldn't find a proof of this, but we have checked it for $n=2$ (four dimensions) and $n=3$ (six dimensions) explicitly, and the case is quite convincing.
This conjecture guarantees that the first $(n-2)$'th expansions in $\Delta_i/|\vec Q_i|$ of (\ref{loop3}) after summing over $i=1,\ldots,n$ vanish, and the non-vanishing result first appears 
in the $(n-1)$'th order as
\bear
&&{(-1)\over (2\pi)^n}{(2n-3)!\over (n-1)!(n-2)!}\,\,(p_1)_{\mu_1}\cdots (p_{n-1})_{\mu_{n-1}}\epsilon^{\mu 0\mu_1\nu_1\cdots\mu_{n-1}\nu_{n-1}}\int^1_0 \prod_{j=1}^{n-1}dx_j\,\,\delta\left(1-\sum_{j=1}^{n-1}x_j\right)\nonumber\\&\times&\int dk^0\int d|\vec k|\,\,\left(\delta(k^0-|\vec k|)+\delta(k^0+|\vec k|)\right)\left({1\over 2}-n_+(k^0)\right)\left(\Delta_i^{n-1}\over|\vec Q_i|^{2n-2}\right)\nonumber\\
&\times&(-1)^i \int_{-1}^1 d\cos\theta {(1-\cos^2\theta)^{n-2}\over (\cos\theta-i k^0\epsilon)^{2n-2}}+ {\cal O}\left(|\vec p|^{n}\right)\,,\label{ithloopn-1}
\eear
where we used the expansion
\be
{1\over (A+x)^{n-1}}=\cdots+(-1)^{n-1} {(2n-3)!\over (n-1)!(n-2)!}{1\over A^{2n-2}}x^{n-1}+\cdots\,.
\ee
The $\cos\theta$ integration can be done as follows. First expand the numerator to obtain
\bear
&&\int_{-1}^1 dx {(1-x^2)^{n-2}\over (x-ik^0\epsilon)^{2n-2}}=\sum_{l=0}^{n-2}(-1)^l {_{n-2}C_l} \int^1_{-1} dx \,{1\over (x-ik^0\epsilon)^{2n-2l-2}}\nonumber\\&=&(-2)\sum_{l=0}^{n-2}{(-1)^l \over (2n-2l-3)}{_{n-2}C_{l}}=(-2)(-1)^n\sum_{l=0}^{n-2}{(-1)^l \over (2l+1)}{_{n-2}C_{l}}\,,
\eear
where in the last equality we change the summation variable $l\to (n-2)-l$.
We now use the identity
\be
\sum_{l=0}^m {(-1)^l \over (2l+1)} {_m C_l}={2^{2m}(m!)^2\over (2m+1)!}\,.\ee
To prove this, start from
\be
\int^1_0 dx \,\,(1-x^2)^m=\sum_{l=0}^m (-1)^l {_m C_l}\int^1_0 dx \,\,x^{2l}=\sum_{l=0}^m {(-1)^l \over (2l+1)} {_m C_l}\,,
\ee
and the left hand side can be computed using the beta function to get the identity proved. Using this identity, the $\cos\theta$ integration finally gives
\be
\int_{-1}^1 dx {(1-x^2)^{n-2}\over (x-ik^0\epsilon)^{2n-2}}=(-2)(-1)^n {2^{2n-4}[(n-2)!]^2\over (2n-3)!}\,.\label{cosint}
\ee
We now conjecture the following result for the Feynman parameter integration in (\ref{ithloopn-1}) after summing over $i=1,\ldots,n$,
\be
({\rm Conjecture}): \,\,\int^1_0 \prod_{j=1}^{n-1}dx_j\,\,\delta\left(1-\sum_{j=1}^{n-1}x_j\right) \sum_{i=1}^n (-1)^i \left(\Delta_i^{n-1}\over |\vec Q_i|^{2n-2}\right)
=-{1\over (n-2)! \,\,2^{2n-3}}\,.\label{conj2}
\ee
We have checked this formula up to $n=5$ (ten dimensions), which is quite non-trivial and convincing. Note that the result doesn't depend on $\vec p_i$'s.
Finally, the $k^0$ and $|\vec k|$ integration in (\ref{ithloopn-1}) gives a simple result
\bear
&&\int dk^0\int d|\vec k|\,\,\left(\delta(k^0-|\vec k|)+\delta(k^0+|\vec k|)\right)\left({1\over 2}-n_+(k^0)\right)\nonumber\\
&=&\int_0^\infty d|\vec k|\,\,\left({1\over 2}-n_+(|\vec k|)+{1\over 2}-n_+(-|\vec k|)\right)=-\mu\,.\label{k0int}
\eear
Collecting (\ref{cosint}), (\ref{conj2}), and (\ref{k0int}), the loop integral (\ref{ithloopn-1}) finally becomes 
\be
{(-1)^n \mu\over (2\pi)^n\,(n-1)!}\,\, (p_1)_{\mu_1}\cdots (p_{n-1})_{\mu_{n-1}}\epsilon^{\mu 0\mu_1\nu_1\cdots\mu_{n-1}\nu_{n-1}}\,,\label{finaresult}
\ee
and this is our final result for the loop integration of (\ref{ithloop}).

Inserting our result into (\ref{ith}) (with $m=n$), we have in momentum space
\bear
J^\mu(p)&=&(-ie)^{n-1}{\mu\over (2\pi)^n (n-1)!}\int{d^{2n}p_1\over (2\pi)^{2n}}\cdots\int{d^{2n}p_{n-1}\over (2\pi)^{2n}} \,\,(2\pi)^{2n}\delta\left(p-p_1-\cdots-p_{n-1}\right)\nonumber\\&\times&
\epsilon^{\mu 0 \mu_1\nu_1\cdots \mu_{n-1}\nu_{n-1}}(p_1)_{\mu_1}\cdots (p_{n-1})_{\mu_{n-1}}A_{\nu_{1}}(p_{1})\cdots A_{\nu_{n-1}}(p_{n-1})\,,
\eear
which becomes in real space,
\bear
J^\mu=(-1)^n e^{n-1}{\mu\over 2^{n-1}(2\pi)^n (n-1)!}
\epsilon^{\mu\nu\mu_1\nu_1\cdots\mu_{n-1}\nu_{n-1}}u_\nu F_{\mu_{1}\nu_{1}}\cdots F_{\mu_{n-1}\nu_{n-1}}\,,
\eear 
where we have introduced the velocity vector of the static fluid $u^\mu=(1,\vec 0)$, and $F_{\mu\nu}=\partial_\mu A_\nu-\partial_\nu A_\mu$ is the field strength. Comparing this with the hydrodynamic prediction in Refs.\cite{Loganayagam:2011mu,Kharzeev:2011ds}, one finds a good agreement, which is an explicit diagrammatic confirmation of the CME in $2n$ dimensions.

\section{Kubo formula and chiral vortical effect in $2n$ dimensions}

The computation in the previous sections can be extended to the response functions to the $0i$ components of the metric perturbations, $\delta g_{0i}$, instead of gauge field perturbations.
At the linear order in $\delta g_{0i}$ in the action, this involves the $0i$-components of the energy-momentum tensors
\be
T^{0i}=-{i\over 4}\psi^\dagger\left(\sigma^0 \overleftrightarrow{\partial}^i\psi+\sigma^i\overleftrightarrow{\partial}^0 \psi\right)\,,\quad \overleftrightarrow{\partial}=\overrightarrow{\partial}-\overleftarrow{\partial}\,.
\ee
These operators couple to the $0i$-components of external metric perturbation $\delta g_{0i}$ in the
action so as to introduce the following additional factor in the Schwinger-Keldysh path integral
\be
\exp\left[i\int_{t_0}^{t_f}\left(T^{0i}_{(1)}-T^{0i}_{(2)}\right)\delta g_{0i}\right]\,.
\ee
The vertex in the Feynman diagrams is generated by
\be
i{\cal L}_{I}=iT^{0i}\delta g_{0i}={1\over 4}\psi^\dagger\left(\sigma^0 \overleftrightarrow{\partial}^i\psi+\sigma^i\overleftrightarrow{\partial}^0 \psi\right)\delta g_{0i}\,,\ee
and let's call this ``Type I'' vertex.
Comparing with the current $J^i=\psi^\dagger \sigma^i \psi$ which couples to $A_i$, 
the structure is similar with the replacements
\be
A_i\to\delta g_{0i}\,,\quad \sigma^i \to -{i\over 4}\left(\sigma^0\overleftrightarrow{\partial}^i+\sigma^i\overleftrightarrow{\partial^0}\right)\,,
\ee
in the vertices, 
so that one can follow similar steps in the previous sections to compute P-odd correlation functions of these energy-momentum vertices.  

The full fermion action in a general metric background is however non-linear in the metric, so there are other terms in the action which are non-linear in the $g_{0i}$ perturbations, and some of them are in fact relevant for our P-odd response functions to the metric perturbations. Following the discussions in Ref.\cite{AlvarezGaume:1983ig}, there are terms containing one $\sigma$ matrix (the lowest term of which is our Type I vertex above) and there are others containing three $\sigma$ matrices coming from spin connection terms, and this class of terms are at least quadratic in $\delta g_{\mu\nu}$. By the same reasoning as in Ref.\cite{AlvarezGaume:1983ig}
one can show that for P-odd correlation functions whose $\epsilon$ tensor emerges from the right number of $\sigma$ and $\bar\sigma$ matrices (that is $2n$) in the numerator, we only need to consider the precisely two types of vertices: the leading Type I vertex with one $\sigma$ matrix and
the leading quadratic vertex containing three $\sigma$ matrices,
\be
i{\cal L}_{II}=-{1\over 16}\left(\psi^\dagger \sigma^{[i}\bar\sigma^\mu\sigma^{j]}\psi\right)\left(\delta g_{0i}\partial_\mu\delta g_{0j}\right)\,,
\ee
where $[i\mu j]=(1/6)(i\mu j \pm {\rm permutations})$ is the anti-symmetrization. We will call this the ``Type II'' vertex.

Let's consider the diagrams for the expectation value of the current $J^\mu(p)$ in response to the $s$ ($s=1,\ldots,(n-1)$) number of $\delta g_{0i}$'s and $t=(n-1)-s$ number of $A_i$'s. We generally have diagrams with $n_1$ number of Type I vertices, $n_2$ Type II vertices, and $t=(n-1)-s$ number of the usual $ie J^i A_i=ie(\psi^\dagger\sigma^i\psi) A_i$ vertices, with a condition $n_1+2n_2=s$. We will compute all these diagrams, and as a first step let's consider the simplest case of $n_2=0$, that is, the
diagrams with only Type I and current vertices without Type II.
They correspond
to replacing $s$ number of current vertices in the previous diagrams with the Type I vertices, and there are ${_{(n-1)}}C_s=(n-1)!/s!(n-1-s)!$ ways of doing it for each $n$ diagrams in the previous section. 
One can easily find that the anticipated P-odd structure of the result in terms of $\epsilon$-tensor
and the external momenta $\vec p_i$'s 
\bear
&\sim& \int{d^{2n}p_1\over (2\pi)^{2n}}\cdots\int{d^{2n}p_{n-1}\over (2\pi)^{2n}} \,\,(2\pi)^{2n}\delta\left(p-p_1-\cdots-p_{n-1}\right) \nonumber\\&\times&\delta g_{0\nu_1}(p_1)\cdots\delta g_{0\nu_s}(p_s)A_{\nu_{s+1}}(p_{s+1})\cdots A_{\nu_{n-1}}(p_{n-1})\nonumber\\&\times&
(p_1)_{\mu_1}\cdots (p_{n-1})_{\mu_{n-1}}\epsilon^{\mu 0\mu_1\nu_1\cdots\mu_{n-1}\nu_{n-1}}\,,
\eear
does not care how these $s$ number of Type I vertices are distributed in the given diagram, so the factor ${_{(n-1)}}C_s=(n-1)!/s!(n-1-s)!$ can simply be multiplied to the result from a single choice of 
the positions of the Type I vertices. Let's then consider the $n$ diagrams as in the previous section where the first $s$ vertices along the arrow directions are replaced by Type I vertices with $\delta g_{0\nu_1}(p_1),\ldots,\delta g_{0\nu_s}(p_s)$.
The denominator is identical, and for the P-odd part of the numerator, we have a replacement
of the first $s$ number of $\sigma^i$'s from the vertex insertions with
\be
\sigma^{i}\to -{i\over 4}\left(\sigma^0\overleftrightarrow{\partial}^i+\sigma^i\overleftrightarrow{\partial^0}\right)\,.
\ee
Each replaced vertex has two pieces: the first one with $\sigma^0$ and the second with $\sigma^i$.
In computing the $\sigma$ matrix trace to get a P-odd $\epsilon$ tensor structure, it is clear that 
one cannot have the first piece appearing twice since that would bring $\sigma^0$ twice in (\ref{sigmatrace}). Therefore the first piece can be chosen at most once. We therefore divide the diagrams into two cases: the Case A where the first piece with $\sigma^0$ never appears, and the Case B where the first piece with $\sigma^0$ appears precisely once. 

{\it Case A:}

Let's first compute the contributions where the first piece is never chosen and all vertex replacement
is simply
\be
\sigma^i \to -{i\over 4}\sigma^i\overleftrightarrow{\partial^0}\,.
\ee
The matrix structure is precisely the same, and in momentum space the presence of the extra $-{i/4}\overleftrightarrow{\partial^0}$ factor
gives ${1/4}$ times the sum of the frequencies of the incoming and out-going momenta.
Since we are considering the limit $p^0_i=0$, the incoming and outgoing frequencies for each vertex is simply $k^0$ of the loop momentum $k^\mu$, so that the factor $-{i/4}\overleftrightarrow{\partial^0}$ simply gives rise to an additional factor $1/4\times (2k^0)=(1/2) k^0$ in the loop integration, compared to the loop integration in the previous section. Since there are $s$ number of them, and including the combinatoric factor  ${_{(n-1)}}C_s=(n-1)!/s!(n-1-s)!$ mentioned in the above, the total contribution
is $(1/2)^s(k^0)^s(n-1)!/s!(n-1-s)! $ times of the expression in the previous section before performing the loop integration.
Since the only modification in the loop integral is the additional $(k^0)^s$, one can simply borrow
the result from the previous section where the previous integral in (\ref{k0int})
\bear
&&\int dk^0\int d|\vec k|\,\,\left(\delta(k^0-|\vec k|)+\delta(k^0+|\vec k|)\right)\left({1\over 2}-n_+(k^0)\right)\nonumber\\
&=&\int_0^\infty d|\vec k|\,\,\left({1\over 2}-n_+(|\vec k|)+{1\over 2}-n_+(-|\vec k|)\right)=-\mu\,,\label{k0int2}
\eear
is now modified by
\bear
&&{1\over 2^s}{(n-1)!\over s! (n-1-s)!}\int dk^0\int d|\vec k|\,\,(k^0)^s\left(\delta(k^0-|\vec k|)+\delta(k^0+|\vec k|)\right)\left({1\over 2}-n_+(k^0)\right)\nonumber\\
&=&{1\over 2^s}{(n-1)!\over s! (n-1-s)!}\int_0^\infty d|\vec k|\,\,|\vec k|^s\left({1\over 2}-n_+(|\vec k|)+(-1)^s\left({1\over 2}-n_+(-|\vec k|)\right)\right)\nonumber\\
&=&{1\over 2^s}{(n-1)!\over s! (n-1-s)!}\int_0^\infty d|\vec k|\,\,|\vec k|^s\left({1\over 2}-n_+(|\vec k|)-(-1)^s\left({1\over 2}-n_-(|\vec k|)\right)\right)\nonumber\\
&=&{1\over 2^s}{(n-1)!\over s! (n-1-s)!}\int_0^\infty d|\vec k|\,\,|\vec k|^s\left({1\over 2}\left(1-(-1)^s\right)-\left(n_+(|\vec k|)-(-1)^s n_-(|\vec k|)\right)\right)\,,\nonumber\\\label{k0int3}
\eear
where we have used the identity
\be
{1\over 2}-n_+(-|\vec k|)=-\left({1\over 2}-n_-(|\vec k|)\right)\,.
\ee
Since in the vacuum we have $n_\pm(|\vec k|)=0$, the first constant piece in the integrand
is the vacuum contribution which is divergent polynomially for odd $s$. In a properly regularized theory, for example by a Pauli-Villars regularization which preserves Lorentz symmetry, 
the regularized finite vacuum result must be Lorentz invariant. However, one can easily see that there is no possible Lorentz symmetric expression that reduces to our expression for our choices for the polarizations, and this means that the regularized vacuum result must vanish identically, so that 
we don't need to introduce renormalized couplings and the renormalized vacuum result is unambiguously zero. Therefore we can ignore
the first piece, so that the final result is a replacement of $-\mu$ in (\ref{k0int2}) or in (\ref{finaresult}) by
\be
-\mu\to -{1\over 2^s}{(n-1)!\over s! (n-1-s)!}\int_0^\infty d|\vec k|\,\,|\vec k|^s\left(n_+(|\vec k|)-(-1)^s n_-(|\vec k|)\right)\,.\label{k0int4}
\ee

{\it Case B:}

We next consider the case where only one replaced vertex among $s$ replaced vertices
has the $\sigma^0$ piece
\be
\sigma^i\to -{i\over 4}\sigma^0\overleftrightarrow{\partial}^i\,,
\ee while the rest $(s-1)$ vertices has the second piece as before
\be
\sigma^i \to -{i\over 4}\sigma^i\overleftrightarrow{\partial^0}={1\over 2}k^0\sigma^i\,.
\ee
There are $s$ number of choices and one can easily find that they all give the same final result, so
let's consider the case where the first vertex along the arrow direction has $-(i/4)\sigma^0\overleftrightarrow{\partial}^i$ while the next $(s-1)$ vertices have $(1/2)k^0\sigma^i$.
The computation of this case is more subtle, but it does contribute to the expected P-odd result.

Including the combinatoric factor, the numerator becomes
\bear
&&(-i)^{n-1}{(n-1)!\over (s-1)!(n-1-s)!}{1\over 2^s}(k^0)^{s-1}\left(k-{p_1\over 2}-p_2-\cdots-p_{i-1}\right)^{\nu_1}\nonumber\\&\times& {\rm tr}\bigg[\sigma^\mu\left((k+p_i+\cdots+p_{n-1})\cdot\bar\sigma\right)\sigma^{\nu_{n-1}}\cdots \left((k+p_i)\cdot\bar\sigma\right)\sigma^{\nu_i}\left(k\cdot\bar\sigma\right)\\&\times&\sigma^{\nu_{i-1}}\left((k-p_{i-1})\cdot\bar\sigma \right)\cdots
\sigma^{\nu_2}\left((k-p_2-\cdots-p_{i-1})\cdot\bar\sigma\right)\sigma^0\left((k-p_1-\cdots-p_{i-1})\cdot\bar\sigma\right)\bigg]\,,\nonumber
\eear
where the meaning of indices $\nu_1,\ldots,\nu_{n-1}$ is that we have the perturbations of
\be\delta g_{0 \nu_1}(p_1),\ldots,\delta g_{0\nu_s}(p_s), A_{\nu_{s+1}}(p_{s+1}),\ldots, A_{\nu_{n-1}}(p_{n-1})\,,
\ee
which is obtained by replacing the first $s$ gauge fields with the metric perturbations in the expression for the $J^\mu(p)$ in (\ref{ith}) with $m=n$ or in the Figure \ref{fig4}.
Performing the trace and extracting the P-odd part gives after some algebra,
\bear
&&(-1)2^{n-1}{(n-1)!\over (s-1)! (n-1-s)!}{1\over 2^s}(k^0)^{s-1}\left(k-{p_1\over 2}-p_2-\cdots-p_{i-1}\right)^{\nu_1}\nonumber\\
&\times&  k_\nu (p_1)_{\mu_1}\cdots (p_{n-1})_{\mu_{n-1}}\epsilon^{\mu 0 \mu_1\nu \mu_2\nu_2\cdots \mu_{n-1}\nu_{n-1}}\,,\label{numenew1}
\eear
which is similar to the previous form (\ref{nume1}) with a few differences. Since $0$-index appears in
the $\epsilon$ tensor, all other indices must be spatial. Especially, we have either a single $k_\nu$ vector or a double vector $k_\nu k^{\nu_1}$ structure that have to be integrated in the loop integral over $k^\mu$. After the same manipulation for the denominator using the Feynman parametrization, the loop integration over the $(2n-1)$ dimensional spatial vector $\vec k$ will be proportional to
\be
\int {d^{2n-1}\vec k\over (2\pi)^{2n-1}}\,\,{k_\nu\over  \left[\vec k\cdot\vec Q_i+\Delta_i-ik^0\epsilon\right]^{n-1}}\sim (\vec Q_i)_\nu\,,
\ee
for the single vector structure, and
\be
\int {d^{2n-1}\vec k\over (2\pi)^{2n-1}}\,\,{k_\nu k^{\nu_1}\over  \left[\vec k\cdot\vec Q_i+\Delta_i-ik^0\epsilon\right]^{n-1}}\sim C_1\,\, (\vec Q_i)_\nu(\vec Q_i)^{\nu_1}+C_2 \,\,\delta^{\nu_1}_\nu \,,
\ee
for the double vector structure by rotational symmetry of the integration measure.
Since $\vec Q_i$ is a linear combination of $\vec p_i$'s, the single vector structure and the first piece of the double vector structure do not contribute to the final result due to the anti-symmetric nature of the $\epsilon$ tensor in (\ref{numenew1}). Therefore, only the second piece in the double vector structure proportional to $\delta^{\nu_1}_\nu$ contributes, and for this purpose we can simply replace
\be
k_\nu k^{\nu_1}\to \delta^{\nu_1}_\nu \cdot{|\vec k_\perp|^2\over (2n-2)}=\delta^{\nu_1}_\nu \cdot{|\vec k|^2\sin^2\theta\over (2n-2)}\,,
\ee
where $\vec k_\perp$ is the component of $\vec k$ which is perpendicular to $\vec Q_i$, and $\theta$ is the angle we introduce in the previous section between $\vec k$ and $\vec Q_i$. The number $(2n-2)$ in the denominator is the number of dimensions of $\vec k_\perp$ that we are averaging over. With all these, our numerator finally becomes almost identical to the previous result (\ref{nume1}) (with $k_\nu=k_0=-k^0$), except the additional factor
\be
{(n-1)!\over (s-1)!(n-1-s)!}{1\over 2^s}{(k^0)^{s-2}|\vec k|^2\sin^2\theta\over (2n-2)}=
{(n-1)!\over (s-1)!(n-1-s)!}{1\over 2^s}{(k^0)^{s}\sin^2\theta\over (2n-2)}\,,\label{additional}
\ee
where we used $|\vec k|^2=(k^0)^2$ due to the delta function structure $\delta(k^0\pm|\vec k|)$ in the rest of the integrand.
As in the {\it Case A} we have an extra $(k^0)^s$ factor, and the presence of $\sin^2\theta$ now modifies the previous angular integration (\ref{cosint})
\be
\int_{-1}^1 dx {(1-x^2)^{n-2}\over (x-ik^0\epsilon)^{2n-2}}=(-2)(-1)^n {2^{2n-4}[(n-2)!]^2\over (2n-3)!}\,,\label{cosint2}
\ee
to a new one
\bear
\int_{-1}^1 dx {(1-x^2)^{n-1}\over (x-ik^0\epsilon)^{2n-2}}&=&(-2)\sum_{l=0}^{n-1}(-1)^l{{_{n-1}C_l}\over (2n-2l-3)}=(-2)(-1)^{n-1}\sum_{l=0}^{n-1}(-1)^l {{_{n-1}C_l}\over (2l-1)}\nonumber\\ &=&(-2)(-1)^n {2^{2n-3}(n-1)! (n-2)!\over(2n-3)!}\,,\label{cosintnew1}
\eear
where in the last line we used a combinatoric identity
\be
\sum_{l=0}^n (-1)^l {{_{n}C_l}\over (l+r)}={n! (r-1)!\over (n+r)!}\,,\label{combid}
\ee
which can be proved by integrating
$
\int^1_0 dx\,\,x^{r-1}(1-x)^n
$ using the beta function.
Comparing (\ref{cosint2}) and (\ref{cosintnew1}), we see that one has an extra factor of $2(n-1)$ from the $\sin^2\theta$ term in the angular integration.
Inserting this to (\ref{additional}), we conclude that the {\it Case B} diagrams give the contribution which is the same to the previous section result (\ref{finaresult})
with a modification
\be -\mu\to  -{1\over 2^s}{(n-1)!\over (s-1)! (n-1-s)!}\int_0^\infty d|\vec k|\,\,|\vec k|^s\left(n_+(|\vec k|)-(-1)^s n_-(|\vec k|)\right)\,.\label{k0int5}
\ee

Note that {\it Case B} result is precisely $s$ times of the {\it Case A} result, so that their sum, which is the final result of the loop integration for the $s$ number of $\delta g_{0i}$ insertions, is (\ref{finaresult}) times
\be
{1\over 2^s}{(n-1)!\over s!(n-1-s)!} (s+1)\,,
\ee
with a replacement
\be \mu\to  \int_0^\infty d|\vec k|\,\,|\vec k|^s\left(n_+(|\vec k|)-(-1)^s n_-(|\vec k|)\right)\,.\label{k0int6}
\ee
We will shortly relate the chiral vortical effect with $s$ number of vorticity insertions to the P-odd response of the current to the $s$ number of $\delta g_{0i}$ perturbations we just computed, after carefully deriving relevant Kubo formula for anomalous transport coefficients in $2n$ dimensions. 
The appearance of the above integration (\ref{k0int6}) in the chiral vortical effect of free chiral fermions was previously predicted in Ref.\cite{Loganayagam:2012pz} using the entropy method of hydrodynamics, and our diagrammatic computation confirms it.
The result of the integration can be found in Ref.\cite{Loganayagam:2012pz}, and it is given in terms of the Bernoulli polynomial as
\bear
&&\int_0^\infty d|\vec k|\,\,|\vec k|^s\left(n_+(|\vec k|)-(-1)^s n_-(|\vec k|)\right)\nonumber\\
&=&{1\over (s+1)}\left(2\pi i\over \beta\right)^{s+1}B_{s+1}\left({1\over 2}+{\beta\mu\over 2\pi i}\right)\sim {1\over (s+1)}\mu^{s+1}+\cdots\,,
\eear
where $\cdots$ involves polynomials of temperature $T$ and $\mu$ which seem to be related to (mixed) gravitational anomalies \cite{Landsteiner:2011cp}. The above formula applies equally well to the $s=0$ case in the previous section.

In summary, the P-odd response of the current $J^\mu(p)$ to the $s$-number of $\delta g_{0i}$ and $(n-1-s)$ number of $A_i$ perturbations coming from the diagrams without any Type II vertices ($n_2=0$) is given by
\bear
&&J^\mu_{(s,t)}(p)\bigg|_{n_2=0}=(-i)^{n-1}\int{d^{2n}p_1\over (2\pi)^{2n}}\cdots\int{d^{2n}p_{n-1}\over (2\pi)^{2n}} \,\,(2\pi)^{2n}\delta\left(p-p_1-\cdots-p_{n-1}\right)\nonumber\\&\times&
\epsilon^{\mu 0 \mu_1\nu_1\cdots \mu_{n-1}\nu_{n-1}}(p_1)_{\mu_1}\cdots (p_{n-1})_{\mu_{n-1}}\delta g_{0\nu_1}(p_1)\cdots \delta g_{0\nu_s}(p_{\nu_s})A_{\nu_{s+1}}(p_{s+1})\cdots A_{\nu_{n-1}}(p_{n-1})\nonumber\\ &\times&
{1\over (2\pi)^n}{1\over 2^s}{1\over s!(n-1-s)!}\left(2\pi i\over\beta\right)^{s+1}B_{s+1}\left({1\over 2}+{\beta\mu\over 2\pi i}\right)\,,\quad s+t=n-1\,.
\eear
In real space, this is equivalent to 
\bear
J^\mu_{(s,t)}\bigg|_{n_2=0}&=&{(-1)^n\over 2^{n-1}(2\pi)^n}{1\over s!(n-1-s)!}\left(2\pi i\over\beta\right)^{s+1}B_{s+1}\left({1\over 2}+{\beta\mu\over 2\pi i}\right)\nonumber\\
&\times&\epsilon^{\mu\nu\mu_1\nu_1\cdots\mu_{n-1}\nu_{n-1}}u_\nu\left(\partial_{\mu_1}\delta g_{0\nu_1}\right)\cdots\left(\partial_{\mu_s}\delta g_{0\nu_s}\right) F_{\mu_{s+1}\nu_{s+1}}\cdots F_{\mu_{n-1}\nu_{n-1}}\,,
\eear 
where we introduce the static velocity vector $u_\nu=-\delta_{\nu 0}$, and $F_{\mu\nu}=\partial_\mu A_\nu -\partial_\nu A_\mu$ is the field strength.

\begin{figure}[t]
	\centering
	\includegraphics[width=10cm]{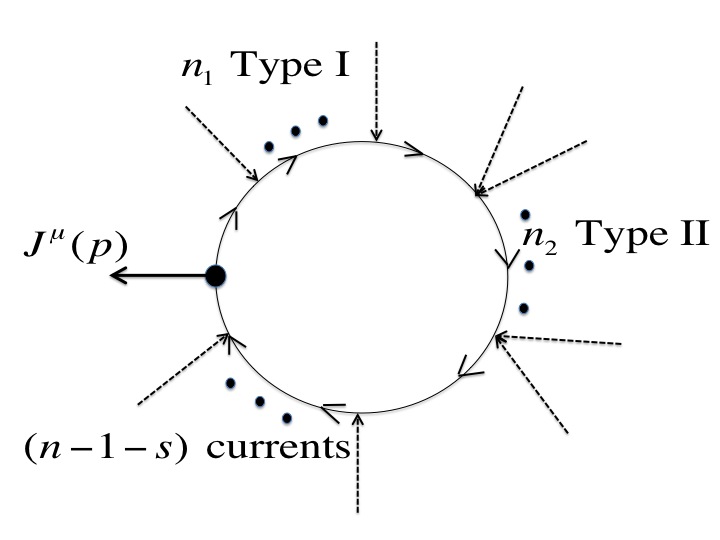}
		\caption{Diagrams for $J^\mu$ with $n_1$ Type I, $n_2$ Type II, and $(n-1-s)$ current vertices. We have $n_1+2n_2=s$, and have to sum over all possible $n_2$ ranging from $0$ to $[s/2]$.\label{fignew1}}
\end{figure}

We now compute the general case of having non-zero $n_2$ number of Type II vertices.
The computation is more or less similar to what we have presented before, except a few minor algebraic differences we will explain in detail.
First, there is an overall combinatoric factor of choosing the positions of Type I and II vertices,
\be
{(n-1-s+n_1+n_2)!\over n_1! n_2! (n-1-s)!}={(n-1-n_2)!\over n_1! n_2! (n-1-s)!}\,,\label{newcomb}
\ee
where we have used $n_1+2n_2=s$.
Since the diagrams with different positions all give the same P-odd result due to $\epsilon$ tensor structure, let's choose $n_1$ Type I vertices to appear first, then $n_2$ Type II, and finally $(n-1-s)$ current vertices, along the arrow direction starting from the current insertion $J^\mu(p)$ as in Figure \ref{fignew1}.
The Feynman rule for the Type II vertex in momentum space is simple: for $\delta g_{0i}(p_1)$ and $\delta g_{0j}(p_2)$ attached to the vertex, one has a vertex insertion
\be
-{i\over 16} (p_1)_\mu\sigma^{[j}\bar\sigma^\mu\sigma^{i]}\,.
\ee
Since we will have an anti-symmetrization for $(i,\mu,j)$ in the final P-odd result by the $\epsilon$ tensor contraction after performing $\sigma$ matrix trace, it is perfectly fine to remove the anti-symmetrization in the above vertex for our computation of P-odd part for simplicity, so that we will use the simpler version in the following,
\be
-{i\over 16} (p_1)_\mu\sigma^{j}\bar\sigma^\mu\sigma^{i}=-{i\over 16} \sigma^{j}(p_1\cdot\bar\sigma)\sigma^{i}\,.
\ee
Comparing this structure with the usual two separate adjacent current insertions with $A_i(p_1)$ and $A_j(p_2)$, now with additional propagator of momentum $p$ between them, 
\be
(ie)^2\sigma^j {-i(p\cdot\bar\sigma)\over -p^2\pm ip^0\epsilon} \sigma^i\,,
\ee
we see that the numerator structure is almost identical. An inspection of the momentum flow in the diagram such as in Figure \ref{fignew1} easily shows that the P-odd part of the numerator is in fact identical to the case with current insertions instead, except
additional numeric factor of $-1/16$ for each Type II. What is non-trivial is that the number of denominators from the propagators
is now reduced from $n$ to $n-n_2\equiv \tilde n$.

Regarding the Type I vertices, the previous classification in terms of the number of $-(i/4)\sigma^0\overleftrightarrow{\partial}^i$ vertices applies here as well, so we have either Case A or Case B. Let's first consider Case A where all Type I vertices are $-(i/4)\sigma^i\overleftrightarrow{\partial}^0$. The above discussion leads to that
the numerator trace gives the result which is
\be
\left(k^0\over 2\right)^{n_1}\times \left(-{1\over 16}\right)^{n_2}
\ee
times of the pure current insertion case (\ref{nume1}).
What is more involved is the angular integration of the denominator since the number of propagators in the denominator integral is reduced by $n_2$.
The integral measure
\be
\int{d^{2n}k\over (2\pi)^{2n}}
={1\over 2^{2n-1}\pi^{n+1}(n-2)!}\int dk^0\int d|\vec k|\,\,|\vec k|^{2n-2}\int^1_{-1} d\cos\theta \,\,\left(1-\cos^2\theta\right)^{n-2} \,,
\ee is the same, but the integrand is now
\bear
&&\sum_{i=0}^{\tilde n} (-1)^{\tilde n-i}\left(-{\pi\over|\vec k|}\right)\left(\delta(k^0-|\vec k|)-\delta(k^0+|\vec k|)\right)\left({1\over 2}-n_+(k^0)\right)\nonumber\\
&\times& (\tilde n -2)!\int_0^1 \prod_{j=1}^{\tilde n-1}dx_j\,\,{\delta\left(1-\sum_j x_j\right)\over \left[|\vec k||\vec {\tilde Q}_i|\cos\theta+\tilde \Delta_i-ik^0\epsilon\right]^{\tilde n-1}}\,,
\eear
with appropriate $(\vec{\tilde Q}_i,\tilde \Delta_i)$ and $\tilde n=n-n_2$, which is essentially the same integrand (\ref{deno3})
for $n_2=0$ case before, but with the replacement $n\to \tilde n$.
Since our previous conjectures (\ref{conj1}) and (\ref{conj2}) are for any $n$ for any momenta $\vec p_i$, they still can be applied to our case with the replacement $n\to\tilde n$. The loop integral then becomes after some algebra (including $k_0(k^0)^{n_1}=-(k^0)^{n_1+1}$ from the numerator),
\bear
&&\int dk^0 \int d|\vec k|\,\,\left(\delta(k^0-|\vec k|)+\delta(k^0+|\vec k|)\right)\left({1\over 2}-n_+(k^0)\right)(k^0)^{n_1+2n_2}\nonumber\\
&\times&{(2\tilde n-3)!\over 2^{2n+2\tilde n-4}\pi^n (n-2)! (\tilde n-1)!(\tilde n-2)!}\times \int^1_{-1}d\cos\theta\,\,{(1-\cos^2\theta)^{n-2}\over (\cos\theta -ik^0\epsilon)^{2\tilde n-2}}\,,
\eear
where we used $|\vec k|^{2n_2}=(k^0)^{2n_2}$ from the delta function piece.
The angular integral can be done as before using the identity (\ref{combid}),
\be
\int^1_{-1}dx\,\,{(1-x^2)^{n-2}\over (x -ik^0\epsilon)^{2\tilde n-2}}=(-1)^{\tilde n-1}2^{2n-4}{(n-2)!(\tilde n-2)! (n-\tilde n-1)!\over (2\tilde n-3)!(2n-2\tilde n-1)!}\,,\label{xintnew1}
\ee
so that the integral finally becomes
\be
{(-1)^{\tilde n-1}(n_2-1)!\over 2^{2\tilde n} \pi^{n} (\tilde n-1)!(2n_2-1)!}\int dk^0 \int d|\vec k|\,\,(k^0)^{s}\left(\delta(k^0-|\vec k|)+\delta(k^0+|\vec k|)\right)\left({1\over 2}-n_+(k^0)\right)\,,
\ee
where we used $n_1+2n_2=s$ and $n-\tilde n=n_2$.
We see that the resulting $(k^0,|\vec k|)$ integral is what we have seen before in (\ref{k0int6}), leading to the same parametric dependence on $(T,\mu)$.
Combining the remaining factors $(1/2)^{n_1}(-1/16)^{n_2}$ from the numerator, and including the combinatoric factor (\ref{newcomb}), the final result after some algebra is the same with the pure current insertion case (\ref{finaresult}) with the replacement
\be
\mu\to {1\over 2^s}{(n-1)!(n_2-1)!\over 2(2n_2-1)!n_1! n_2! (n-1-s)!}\int_0^\infty d|\vec k|\,\,|\vec k|^s\left(n_+(|\vec k|)-(-1)^s n_-(|\vec k|)\right)\,.\label{newk0intA}
\ee
This is a generalization of (\ref{k0int4}) to a non-zero $n_2$, and one can check that it indeed reduces to (\ref{k0int4}) correctly in $n_2=0$ limit.

The Case B where one of the Type I vertices has $-(i/4)\sigma^0\overleftrightarrow{\partial}^i$ is also computed similarly as before.
The net result is that one has an additional combinatoric factor $n_1$ from the possible choices of the Type I vertex which has 
$-(i/4)\sigma^0\overleftrightarrow{\partial}^i$, and the angular integration gets an additional factor
\be
{\sin^2\theta\over (2n-2)}={(1-\cos^2\theta)\over (2n-2)}\,,
\ee
so that the angular integration is now modified to
\be
{1\over (2n-2)}\int^1_{-1}dx\,\,{(1-x^2)^{n-1}\over (x-ik^0\epsilon)^{2\tilde n-2}}={(-1)^{\tilde n-1} 2^{2n-2}(n-1)! (\tilde n-2)! (n-\tilde n)!\over (2n-2) (2\tilde n-3)! (2n-2\tilde n+1)!}\,.\label{xintnew2}
\ee
Comparing with the previous angular integration (\ref{xintnew1}), this is $1/(2n-2\tilde n+1)=1/(2n_2+1)$ times of (\ref{xintnew1}).
Combining the additional combinatoric factor $n_1$, this finally concludes that the Case B contribution is $n_1/(2n_2+1)$ times of the Case A, so that the sum of Case A and B, which is the final result, is $(n_1+2n_2+1)/(2n_2+1)=(s+1)/(2n_2+1)$ times of the Case A result.

In summary, the final result for $n_2$ number of Type II vertices insertion is 
the same with the pure current insertion case (\ref{finaresult}) with the replacement
\bear
\mu&\to& {(s+1)\over 2^s}{(n-1)!\over (n-1-s)!}{1\over (2n_2+1)!(s-2n_2)!}\int_0^\infty d|\vec k|\,\,|\vec k|^s\left(n_+(|\vec k|)-(-1)^s n_-(|\vec k|)\right)\nonumber\\
&=&{1\over 2^s}{(n-1)!\over (n-1-s)!}{1\over (2n_2+1)!(s-2n_2)!}\left(2\pi i\over \beta\right)^{s+1}B_{s+1}\left({1\over 2}+{\beta\mu\over 2\pi i}\right)
\,.\label{newtotal}
\eear
What we have to do 
lastly is to sum up all contributions with all possible $n_2$ ranging from $0$ to $[s/2]$. Magically this is doable compactly, using the combinatoric identity
\be
\sum_{n_2=0}^{\left[{s\over 2}\right]}{1\over (2n_2+1)!(s-2n_2)!}={1\over (s+1)!}\sum_{n_2=0}^{\left[{s\over 2}\right]} {_{s+1}}C_{2n_2+1}={2^s\over (s+1)!}\,.\label{n2sum}
\ee

With all these, the final response current in real space with $s$ number of $\delta g_{0i}$ perturbations and $t=n-1-s$ number of gauge fields is 
\bear
J^\mu_{(s,t)}&=&{(-1)^n\over 2^{n-1-s}(2\pi)^n}{1\over (s+1)!(n-1-s)!}\left(2\pi i\over\beta\right)^{s+1}B_{s+1}\left({1\over 2}+{\beta\mu\over 2\pi i}\right)\nonumber\\
&\times&\epsilon^{\mu\nu\mu_1\nu_1\cdots\mu_{n-1}\nu_{n-1}}u_\nu\left(\partial_{\mu_1}\delta g_{0\nu_1}\right)\cdots\left(\partial_{\mu_s}\delta g_{0\nu_s}\right) F_{\mu_{s+1}\nu_{s+1}}\cdots F_{\mu_{n-1}\nu_{n-1}}\,,
\eear 
Following the notation in Ref.\cite{Kharzeev:2011ds}, we define
\be
\Delta B^\mu_{(s,t)}\equiv {1\over n}\epsilon^{\mu\nu\mu_1\nu_1\cdots\mu_{n-1}\nu_{n-1}}u_\nu\left(\partial_{\mu_1}\delta g_{0\nu_1}\right)\cdots\left(\partial_{\mu_s}\delta g_{0\nu_s}\right) F_{\mu_{s+1}\nu_{s+1}}\cdots F_{\mu_{n-1}\nu_{n-1}}\,,\label{DeltaBi}
\ee
where we reserve the notation $B^\mu_{(s,t)}$ for the true chiral vortical current
\be
B^\mu_{(s,t)}\equiv {1\over n}\epsilon^{\mu\nu\mu_1\nu_1\cdots\mu_{n-1}\nu_{n-1}}u_\nu\left(\partial_{\mu_1}u_{\nu_1}\right)\cdots\left(\partial_{\mu_s} u_{\nu_s}\right) F_{\mu_{s+1}\nu_{s+1}}\cdots F_{\mu_{n-1}\nu_{n-1}}\,.\label{Bst}
\ee
Then our results can be written as
\bear
J^\mu&=&\sum_{s+t=n-1}{(-1)^n\over 2^{n-1-s}(2\pi)^n}{n\over (s+1)!(n-1-s)!}\left(2\pi i\over\beta\right)^{s+1}B_{s+1}\left({1\over 2}+{\beta\mu\over 2\pi i}\right)\,\, \Delta B^\mu_{(s,t)}\nonumber\\
&\equiv& \sum_{s+t=n-1} \xi^{\rm AF}_{(s,t)} \,\,\Delta B^\mu_{(s,t)}\,,\label{diares1}
\eear
with the transport coefficients $\xi^{\rm AF}_{(s,t)}$ the meaning of whose superscript AF (Anomaly Frame) will become clear when we discuss the Kubo formula shortly.

Up to now, we have computed the P-odd response of the current $J^\mu(p)$ to the external $\delta g_{0i}$ and $A_i$ perturbations.
It is straightforward to compute the P-odd response of the energy-momentum $T^{0i}$ to the same perturbations.
One caveat is that due to the presence of non-linear terms of metric perturbations in the action, the energy-momentum itself is also modified from its flat space one, $T^{0i}_I=(-i/4)(\sigma^0\overleftrightarrow{\partial}^i +\sigma^i\overleftrightarrow{\partial}^0)$, by additional terms involving $\delta g_{0i}$ explicitly. Since only Type I and Type II terms in the action are relevant in P-odd response functions, we only need to compute the correction coming from Type II term to the energy-momentum tensor for our P-odd response function. This is given by
\be
T^{0i}_{II}={i\over 16}\left((\psi^\dagger\sigma^{[i}\bar\sigma^\mu\sigma^{j]}\psi)(\partial_\mu \delta g_{0j})+\partial_\mu(\psi^\dagger \sigma^{[i}\bar\sigma^\mu\sigma^{j]}\psi \delta g_{0j})\right)\,.
\ee

Let's first consider the contribution from the flat space energy-momentum, $T^{0i}_I$,
by simply replacing the current vertex $\sigma^i$ with the $T^{0i}$, that is, $(-i/4)(\sigma^0\overleftrightarrow{\partial}^i +\sigma^i\overleftrightarrow{\partial}^0)$.
If we insert the second piece ({\it Case A}), there is no change in the matrix trace and we simply get an additional factor of $k^0/2$ compared to the above computation for $J^\mu_{(s,t)}$, so that this gives a contribution
\be
T^{0i}_{(s,t)}\bigg |_A={(-1)^n\over 2^{n-s}(2\pi)^n}{n\over s!(n-1-s)!}{1\over (s+2)}\left(2\pi i\over\beta\right)^{s+2}B_{s+2}\left({1\over 2}+{\beta\mu\over 2\pi i}\right)\,\, \Delta B^i_{(s,t)}\,:{\it Case\,\,A}\label{t0iA}
\ee
On the other hand, the insertion of the first piece ({\it Case B}) can be treated by precisely the same way as before.
Considering $n_1$ Type I, $n_2$ Type II, $(n-1-s)$ current vertices, we have the same combinatoric factor
\be
{(n-1-n_2)!\over n_1! n_2 ! (n-1-s)!}\,,
\ee 
and additional numeric factors
\be
\left({1\over 2}\right)^{n_1+1} \left(-{1\over 16}\right)^{n_2}\,,
\ee
and the angular integration is modified by the additional factor
\be
{\sin^2\theta\over 2(n-1)}\,.
\ee
The total number of $|\vec k|$ in the integration similar to (\ref{newtotal}) is now $s+1$ instead of $s$.
The summation over $n_2$ is done precisely by the same way as in (\ref{n2sum}). The result is
\be
T^{0i}_{(s,t)}\bigg |_B={(-1)^n\over 2^{n-s}(2\pi)^n}{n\over (s+2)!(n-1-s)!}\left(2\pi i\over\beta\right)^{s+2}B_{s+2}\left({1\over 2}+{\beta\mu\over 2\pi i}\right)\,\, \Delta B^i_{(s,t)}\,:{\it Case\,\,B}\label{t0iB}
\ee
which is $1/(s+1)$ times of the Case A result.

\begin{figure}[t]
	\centering
	\includegraphics[width=10cm]{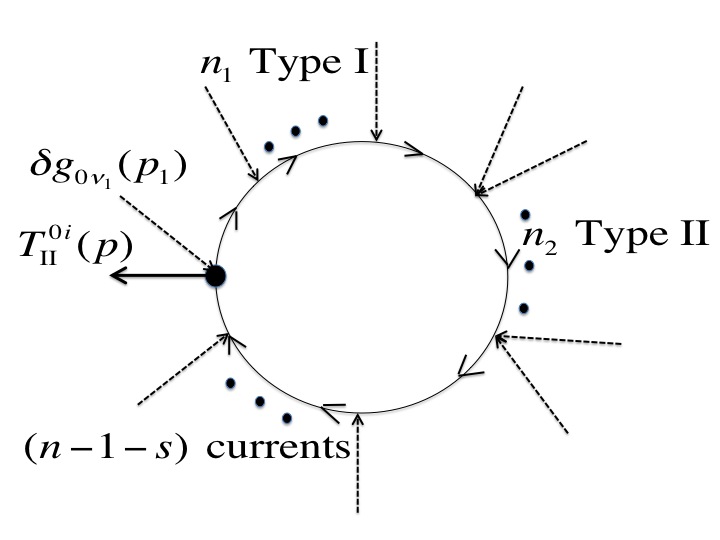}
		\caption{Diagrams for $T^{0i}_{II}$ contribution to the total $T^{0i}$ response function. We have a constraint $n_1+2n_2=(s-1)$, and have to sum over all $n_2$ ranging from $0$ to $[(s-1)/2]$. \label{fignew2}}
\end{figure}
We next compute the contribution coming from the $T^{0i}_{II}$ vertex ({\it Case C}).
Note that since $T^{0i}_{II}$ contains one $\delta g_{0\nu_1}$ already, we should have $(s-1)$ number of additional $\delta g_{0i}$ insertions in the diagrams, so that $n_1+2n_2=(s-1)$ and the number of propagators (denominators) is $n-n_2-1=\tilde n-1$.
See Figure \ref{fignew2} for the diagrams of the {\it Case C}.
Working out the Feynman rule in the momentum space, the $T^{0i}_{II}$ insertion with $\delta g_{0\nu_1}(p_1)$ attached corresponds to
\be
-{1\over 16}\sigma^{[i}\bar\sigma^\mu\sigma^{\nu_1]}\left(p_1+p_{\rm tot}\right)_\mu\delta g_{0\nu_1}(p_1)\,,
\ee
where $p_{\rm tot}$ is the total momentum flowing out from the $T^{0i}_{II}$, which is simply
\be
p_{\rm tot}=p_1+p_2+\cdots + p_{n-1}\,,
\ee in our diagrams. Since $\sigma$ trace will totally anti-symmetrize $(i\mu\nu_1)$ indices anyway, one can safely remove anti-symmetrization in the above vertex. Looking at the final $\sigma$ trace, only the term proportional to $p_1$ survives the total anti-symmetrization (since $p_2, \cdots ,p_{n-1}$ are necessarily contracted with $\epsilon$ tensor already from other parts of propagators), so that the vertex takes a final form
\be
-{1\over 8}\sigma^i (p_1\cdot\bar\sigma)\sigma^{\nu_1}\,.
\ee  
Comparing this with the previous pure current response function which will have the corresponding numerator piece
\be
(ie) \sigma^i i (p_1\cdot\bar\sigma)\sigma^{\nu_1}=-e\sigma^i (p_1\cdot \bar\sigma)\sigma^{\nu_1}\,,
\ee
we see that the effect of $T^{0i}_{II}$ insertion for the numerator is an additional simple factor $1/8$ compared to the pure current
result (\ref{nume1}). 
The only other effects remaining are the modification of the relation $n_1+2n_2=(s-1)$ and the number of denominators $n-n_2-1=\tilde n-1$. Since all previous computations such as (\ref{xintnew1}) and (\ref{xintnew2}) are derived for any $(n_1,n_2,\tilde n)$,
it is straightforward to repeat the previous algebra to compute these diagrams.
In the first case where all $n_1$ Type I vertices are $(-i/4)(\sigma^i\overleftrightarrow{\partial}^0)$, we have combinatoric and numeric factors
\be
{(n-1-s+n_1+n_2)!\over n_1! n_2! (n-1-s)!}\times {1\over 8}\times\left(k^0\over 2\right)^{n_1}\times \left(-{1\over 16}\right)^{n_2}\,,
\ee
and the angular integration involved is (since the number of denominators is now $\tilde n-1$ instead of $\tilde n$)
\be
\int^1_{-1}dx {(1-x^2)^{n-2}\over (x-ik^0\epsilon)^{2\tilde n-4}} = (-1)^{\tilde n-2}{2^{2n-4}(n-2)!(\tilde n-3)!(n-\tilde n)!\over (2\tilde n-5)!(2n-2\tilde n+1)!}\,,\label{xnewint3}
\ee
which is (\ref{xintnew1}) with $\tilde n\to\tilde n-1$. 
In the other case where one Type I vertex is $(-i/4)(\sigma^0\overleftrightarrow{\partial}^i)$ while the rest $(n_1-1)$ are $(-i/4)(\sigma^i\overleftrightarrow{\partial}^0)$, there is an additional combinatoric factor $n_1$ while the angular integration has an extra factor $\sin^2\theta/(2n-2)$, so that it becomes
\be
{1\over 2n-2}\int^1_{-1}dx {(1-x^2)^{n-1}\over (x-ik^0\epsilon)^{2\tilde n-4}}\,,
\ee
which is $1/(2n-2\tilde n+3)=1/(2n_2+3)$ times of (\ref{xnewint3}), so that this second case is $n_1/(2n_2+3)$ times of the first case. Summing these two cases, 
after straightforward algebra, produces the result with $n_2$ number of Type II vertices as
\be
T^{0i}_{(s,t)}\bigg |_{C,n_2}={(-1)^n\over 2^{n}(2\pi)^n}{n\over (n-1-s)!}{2(n_2+1)\over (s-1-2n_2)!(2n_2+3)!}\left(2\pi i\over\beta\right)^{s+2}B_{s+2}\left({1\over 2}+{\beta\mu\over 2\pi i}\right)\,\, \Delta B^i_{(s,t)}\,,\label{t0iCb}
\ee
which has to be summed over $n_2$ ranging from $0$ to $[(s-1)/2]$. Amusingly, this summation can be performed by the combinatoric identity
\be
\sum_{n_2=0}^{[(s-1)/2]} {(n_2+1)\over (s-1-2n_2)!(2n_2+3)!}={s\cdot 2^{s-1}\over (s+2)!}\,,
\ee
so that the final result of the Case C is
\be
T^{0i}_{(s,t)}\bigg |_{C}={(-1)^n\over 2^{n-s}(2\pi)^n}{n\over (n-1-s)!}{s\over (s+2)!}\left(2\pi i\over\beta\right)^{s+2}B_{s+2}\left({1\over 2}+{\beta\mu\over 2\pi i}\right)\,\, \Delta B^i_{(s,t)}\,:{\it Case\,\,C}\label{t0iC}
\ee
Summing over all cases A, B, and C lastly gives the final result for the energy-momentum response
\be
T^{0i}_{(s,t)}={(-1)^n\over 2^{n-1-s}(2\pi)^n}{n\over s!(n-1-s)!}{1\over (s+2)}\left(2\pi i\over\beta\right)^{s+2}B_{s+2}\left({1\over 2}+{\beta\mu\over 2\pi i}\right)\,\, \Delta B^i_{(s,t)}\,.\label{t0ifinal}
\ee
In summary, the total $T^{0i}$ response is
\be
T^{0i}=\sum_{s+t=n-1} \lambda^{\rm AF}_{(s,t)}\,\, \Delta B^i_{(s,t)}\,,\label{diares2}
\ee
with the transport coefficients
\be
\lambda_{(s,t)}^{\rm AF}={(-1)^n\over 2^{n-1-s}(2\pi)^n}{n\over s!(n-1-s)!}{1\over (s+2)}\left(2\pi i\over\beta\right)^{s+2}B_{s+2}\left({1\over 2}+{\beta\mu\over 2\pi i}\right)\,.
\ee
This completes our diagrammatic computations.

We now discuss the Kubo formula for anomalous transport coefficients, generalizing Ref.\cite{Amado:2011zx} in $2n=4$ dimensions to arbitrary dimensions. The basic idea is the following: what we have computed above is the P-odd response
of the current and $T^{0i}$ in the presence of the external gauge field and metric $\delta g_{0i}$ perturbations.
By computing the same response in the framework of hydrodynamics with unknown P-odd anomalous transport coefficients and comparing with what we have computed,
one can determine the P-odd anomalous transport coefficients. Strictly speaking, the free fermion theory we are considering does not have a hydrodynamic regime, so that this
procedure should not be applicable in principle. What has been assumed and also showed in specific cases is that the zero frequency limit of the free theory computation, or equivalently the Euclidean correlation functions,
is not renormalized in the presence of the interactions \cite{Golkar:2012kb} \footnote{See Refs.\cite{Golkar:2012kb,Hou:2012xg,Jensen:2013vta} for the exceptions when the external gauge fields become dynamical.}, so that one may get the correct result even from the free theory computation of the same quantities. Our discussion is based on this expectation extended to $2n$ dimensions. There are also evidences for this in the effective action approach \cite{Banerjee:2012cr,Jensen:2013kka,Jensen:2013rga,Haehl:2013hoa}.

There is an ambiguity in defining the hydrodynamics, which corresponds to the choice of the fluid vector $u^\mu$.
We discuss Kubo formula in two such ``frame'' choices: ``Anomaly frame'' and Landau frame.

{\it Anomaly frame:}

The anomaly frame, which was introduced in Ref.\cite{Loganayagam:2011mu}, is the frame where the anomalous transport effects appearing in the current and energy-momentum constitutive relations
take the simplest form,
\bear
J^\mu&=&\rho u^\mu+\sigma\left(E^\mu-T\,\,\Pi^{\mu\nu}\nabla_\nu\left(\mu\over T\right)\right)+\cdots+\sum_{s+t=n-1}\xi_{(s,t)}^{\rm AF} B^\mu_{(s,t)}+\cdots\,,\nonumber\\
T^{\mu\nu}&=&(\epsilon+p)u^\mu u^\nu +p g^{\mu\nu}-2\eta \sigma^{\mu\nu}+\cdots+\sum_{s+t=n-1}\lambda^{\rm AF}_{(s,t)}\left(u^\mu B^\nu_{(s,t)}+u^\nu B^\mu_{(s,t)}\right)+\cdots\,,\nonumber\\\label{AFcons}
\eear
where $B^\mu_{(s,t)}$ is defined in (\ref{Bst}) above, and $\cdots$ means any lower or higher order terms which are P-even.
The $\eta$ is the shear viscosity\footnote{We ignore the bulk viscosity term $-\zeta \Pi^{\mu\nu}\nabla_\alpha u^\alpha$ since it does not affect our following discussion.}, and
\be
\sigma^{\mu\nu}={1\over 2}\Pi^{\mu\alpha}\Pi^{\nu\beta}\left(\nabla_\alpha u_\beta +\nabla_\beta u_\alpha-{2\over (2n-1)}(\nabla_\gamma u^\gamma)\right)\,,\quad \Pi^{\mu\nu}\equiv g^{\mu\nu}+u^\mu u^\nu\,.
\ee
It is important to emphasize that there are no other P-odd transport effects other than what is shown above appearing in $(n-1)$'th order in derivatives, which is
a major advantage of working in the anomaly frame \cite{Loganayagam:2011mu} \footnote{This frame is also characterized by the absence of anomaly generated entropy flow. We thank Misha Stephanov for pointing this to us.}.

Let's introduce the spatial gauge field $A_i$ and the metric $\delta g_{0i}$ perturbations in the static limit and solve the hydrodynamic equations \footnote{we ignore the anomaly term in $\nabla_\mu J^\mu=0$ since we don't have electric fields.}
\be
\nabla_\mu J^\mu=0\,,\quad \nabla_\mu T^{\mu\nu}=0\,,
\ee
with the above constitutive relations. This means to solve for the perturbations of hydrodynamic degrees of freedom, $(u^\mu, p,\rho)$, induced by the external perturbations $(A_i,
\delta g_{0i})$. Since the hydrodynamics of a given frame choice is a self-contained dynamical system of equations, one expects to find a unique answer with reasonable boundary conditions at infinity. 
We would like to obtain a linearized and leading derivative contribution to $(\delta u^\mu,\delta p,\delta\rho)$ from non-anomalous hydrodynamic response, while we would like to trace the first leading effect from anomaly which appears at $(n-1)$'th order. Since $\delta B^\mu_{(s,t)}$ is already $(n-1)$'th order in terms of $\delta u^\mu$ and $A_i$, it is sufficient to use the leading expressions for $\delta u^\mu$ in computing $\delta B^\mu_{(s,t)}$ for our purpose. However, we may still need to keep $(n-1)$'th order corrections to $(\delta u^\mu,\delta p,\delta\rho)$ coming from anomaly to obtain the correct $(n-1)$'th order corrections to the response of $J^i$ and $T^{0i}$ from anomaly. This point will in fact be important in the Landau frame choice discussed later.

First, from $u^\mu u^\nu g_{\mu\nu}=-1$, we have $\delta u^0=0$. It is easy to derive $\delta \Gamma^\mu_{\mu0}=0$, so that $\nabla_\mu J^\mu=0$ gives $\partial_i\delta J^i=0$. On the other hand,  
\be
\delta J^i=\rho \delta u^i -\sigma T\partial_i \delta\left(\mu\over T\right)+\cdots +\sum_{s+t=n-1}\xi^{\rm AF}_{(s,t)} \delta B^i_{(s,t)}+\cdots\,,\label{deltaji}
\ee
where all quantities without $\delta$ mean those in the unperturbed equilibrium state, and
\be
\delta B^i_{(s,t)}={1\over n}\epsilon^{i\nu\mu_1\nu_1\cdots \mu_{n-1}\nu_{n-1}}u_\nu \left(\partial_{\mu_1}\delta u_{\nu_1}\right)\cdots\left(\partial_{\mu_s}\delta u_{\nu_{s}}\right) F_{\mu_{s+1}\nu_{s+1}}\cdots F_{\mu_{n-1}\nu_{n-1}}\,.
\ee
It is easy to see that $\partial_i \delta B^i_{(s,t)}=0$ due to $\epsilon$-tensor, so we get from $\partial_i\delta J^i=0$,
\be
\rho \partial_i\delta u^i-\sigma T\partial_i\partial_i\delta\left(\mu\over T\right)=0\,,
\ee
up to leading non-trivial order in derivatives, and importantly the leading anomaly induced effect appearing at $(n-1)$'th order  is absent in this equation.
Next, from $\delta \Gamma^0_{ii}=-\partial_i \delta g_{0i}$, the variation of $\nabla_\mu T^{\mu 0}=0$ gives
\be
\partial_i \delta T^{0i}-p \partial_i\delta g_{0i}=0\,,
\ee
while the variation of $T^{0i}$ is
\be
\delta T^{0i}=(\epsilon+p)\delta u^i+p\delta g_{0i}+\cdots+\sum_{s+t=n-1}\lambda^{AF}_{(s,t)}\delta B^i_{(s,t)}+\cdots\,,\label{deltat0i}
\ee
so that we get from $\nabla_\mu T^{\mu 0}=0$,
\be
\partial_i\delta u^i=0\,,\label{uii}
\ee
and again the leading anomaly contribution is absent in this equation. Finally, the variation of the equation $\nabla_\mu T^{\mu j}=0$ can be shown to become
\be
\partial_i \delta T^{ij}=0\,,
\ee
whereas
\be
\delta T^{ij}=(\delta p) \delta^{ij}-2\eta\delta\sigma^{ij}=(\delta p)\delta^{ij}-\eta\left(\partial^i\delta u^j+\partial^j\delta u^i-{2\over (2n-1)}\delta^{ij}(\partial_k\delta u^k)\right)\,,
\ee
which leads to
\be
\partial_j\delta p-\eta\left(\partial_i\partial^i\delta u^j +{(2n-3)\over (2n-1)}\partial_j\left(\partial_i\delta u^i\right)\right)=\partial_j\delta p-\eta\partial_i\partial^i\delta u^j=0\,,
\ee
where we used (\ref{uii}). Taking $\partial_j$ to the above and using (\ref{uii}) again gives
\be
\partial_j\partial^j\delta p=0\,.
\ee
The ellipticity of Laplace equation gives then $\delta p=0$, which subsequently implies $\delta u^i=0$ and $\delta \rho=0$ at leading order. These results will in general be modified if we include higher order P-even corrections, but
what should be emphasized is the absence of leading anomaly contribution at $(n-1)$'th order in the above results of $(\delta u^\mu,\delta p,\delta \rho)$.
Inserting these results to (\ref{deltaji}) and (\ref{deltat0i}), and using the identity
\be
\delta u_i=\delta g_{0i}+\delta u^i\,,
\ee
so that we have
\be
\delta B^i_{(s,t)}=\Delta B^i_{(s,t)}\,,
\ee where $\Delta B^i_{(s,t)}$ is defined previously in (\ref{DeltaBi}),  
one finally concludes that the leading P-odd response of the current $J^i$ and $T^{0i}$ at $(n-1)$'th order is
given by
\be
J^i_{\rm P-odd}=\sum_{s+t=n-1}\xi^{\rm AF}_{(s,t)} \Delta B^i_{(s,t)}\,,\quad T^{0i}_{\rm P-odd}=\sum_{s+t=n-1}\lambda^{\rm AF}_{(s,t)} \Delta B^i_{(s,t)}\,.
\ee
Comparing this with our diagrammatic computation (\ref{diares1}) and (\ref{diares2}), we see that what we called $(\xi^{\rm AF}_{(s,t)},\lambda^{\rm AF}_{(s,t)})$ in (\ref{diares1}) and (\ref{diares2}) indeed coincide with the anomalous transport coefficients in the anomaly frame appearing in the constitutive relations (\ref{AFcons}).

{\it Landau frame:}

The discussion in the Landau frame is slightly more complicated. 
As shown in Ref.\cite{Kharzeev:2011ds}, the leading $(n-1)$'th order effect from anomaly appears only in the current constitutive relation
\be
J^\mu=\rho u^\mu+\sigma\left(E^\mu-T\,\,\Pi^{\mu\nu}\nabla_\nu\left(\mu\over T\right)\right)+\cdots+\sum_{s+t=n-1}\xi_{(s,t)}^{\rm LF} B^\mu_{(s,t)}+\cdots\,,\label{LFcons1}
\ee
whereas the energy-momentum may get contributions starting at one order higher, that is, at $n$'th order in derivative,
\be
T^{\mu\nu}=(\epsilon+p)u^\mu u^\nu +p g^{\mu\nu}-2\eta \sigma^{\mu\nu}+\cdots+{\eta \over \epsilon+p}\sum_{s+t=n-1}\lambda^{\rm LF}_{(s,t)}\Pi^{\mu\alpha}\Pi^{\nu\beta}\left(\nabla_{\alpha}B_{(s,t)\beta}+\nabla_{\beta} B_{(s,t)\alpha}\right)+\cdots\,,\label{LFcons2}
\ee
where we showed only one possible $n$'th order contributions since they turn out to be relevant, giving rise to a $(n-1)$'th order correction to $\delta u^i$ coming from anomaly.
The previous discussion up to (\ref{uii}) is the same, leading to $\partial_i \delta u^i=0$.
For $\partial_i \delta T^{ij}=0$, we now instead have
\bear
\delta T^{ij}&=&(\delta p)\delta^{ij}-\eta\left(\partial^i\delta u^j+\partial^j\delta u^i-{2\over (2n-1)}\delta^{ij}(\partial_k\delta u^k)\right)\nonumber\\&+&
{\eta\over\epsilon+p}\sum_{s+t=n-1}\lambda^{\rm LF}_{(s,t)}\left(\partial^i\delta B^j_{(s,t)}+\partial^j\delta B^i_{(s,t)}\right)\,,
\eear
which gives the equation
\be
\partial^j\delta p-\eta\partial_i\partial^i \delta u^j+{\eta\over\epsilon+p}\sum_{s+t=n-1}\lambda^{\rm LF}_{(s,t)}\partial_i\partial^i\delta B^j_{(s,t)}=0\,,
\ee 
where we used $\partial_i\delta u^i=0$ and $\partial_i\delta B^i_{(s,t)}=0$. Taking $\partial_j$ to the above then gives $\partial_i\partial^i\delta p=0$, so that $\delta p=0$, and we have
\be
-\partial_i\partial^i \delta u^j+{1\over\epsilon+p}\sum_{s+t=n-1}\lambda^{\rm LF}_{(s,t)}\partial_i\partial^i\delta B^j_{(s,t)}=0\,,
\ee
which finally gives
\be
\delta u^i=0+\cdots+{1\over\epsilon+p}\sum_{s+t=n-1}\lambda^{\rm LF}_{(s,t)}\delta B^i_{(s,t)}\,,
\ee
where $\cdots$ means all possible P-even contributions beyond leading order, but the main point is that we have identified the leading $(n-1)$'th order effect from anomaly to $(\delta u^i,\delta p,\delta\rho)$ unambiguously. Inserting these to (\ref{LFcons1}) and (\ref{LFcons2}) produces the leading effects from anomaly at $(n-1)$'th order as
\be
J^i_{\rm P-odd}=\sum_{s+t=n-1}\left(\xi^{\rm LF}_{(s,t)}+{\rho\over\epsilon+p}\lambda^{\rm LF}_{(s,t)}\right)\Delta B^i_{(s,t)}\,,\quad T^{0i}_{\rm P-odd}=\sum_{s+t=n-1}\lambda^{\rm LF}_{(s,t)} \Delta B^i_{(s,t)}\,.
\ee
Comparing this with our diagrammatic computation (\ref{diares1}) and (\ref{diares2}), we conclude that
\be
\xi^{\rm LF}_{(s,t)}=\xi^{\rm AF}_{(s,t)}-{\rho\over\epsilon+p}\lambda^{\rm AF}_{(s,t)}\,,\quad \lambda^{\rm LF}_{(s,t)}=\lambda^{\rm AF}_{(s,t)}\,.
\ee
Note that the transport coefficients $\lambda^{\rm LF}_{(s,t)}$ in the Landau frame appear as $n$'th order transport coefficients naively.

\section{Discussion}

Comparing our results with the predictions from hydrodynamics in Refs.\cite{Loganayagam:2011mu,Kharzeev:2011ds}, we find that our results for $\xi^{\rm AF}_{(s,t)}$ and $\lambda^{\rm AF}_{(s,t)}$ remarkably agree with the hydrodynamics results. Our results for the Landau frame transport coefficients $\xi^{\rm LF}_{(s,t)}$ take the form,
\bear
\xi^{\rm LF}_{(s,t)}&=&\xi^{\rm AF}_{(s,t)}-{\rho\over\epsilon+p}\lambda^{\rm AF}_{(s,t)}\nonumber\\&=&C_n{2^s\over (s+1)!(n-1-s)!}\Bigg(\left(2\pi i\over\beta\right)^{s+1}B_{s+1}\left({1\over 2}+{\beta\mu\over 2\pi i}\right)\nonumber\\&&-{\rho\over \epsilon+p}{(s+1)\over (s+2)}\left(2\pi i\over\beta\right)^{s+2}B_{s+2}\left({1\over 2}+{\beta\mu\over 2\pi i}\right)\Bigg)\,,
\eear
where \be
C_n={(-1)^n n\over 2^{n-1}(2\pi)^n}\,,
\ee
is a constant that depends only on the dimension $2n$.
Using $B_m(x)=x^m+\cdots$, and looking at the terms which contain only $\mu$, neglecting terms involving powers of temperature $T=\beta^{-1}$, we have
\be
\xi^{\rm LF}_{(s,t)}=C_n{2^s\over (s+1)!(n-1-s)!}\left(\mu^{s+1}-{\rho\over \epsilon+p}{(s+1)\over (s+2)}\mu^{s+2}\right)+{\rm powers}\,\,{\rm of}\,\,T\,,
\ee
which agrees with the eq.(3.157) of Ref.\cite{Kharzeev:2011ds} with the identification $\kappa=C_n/(n-1)!$. The correct $s$ dependence should be noted.
Given that we have summed over many diagrams with different topologies, the agreement seems quite non-trivial, and provides an explicit diagrammatic confirmation of the hydrodynamic predictions.

The properties of spinor algebra are periodic in dimensions with a period of 8 dimensions. Correspondingly, the Hamiltonian describing the quantized one particle state naturally realizes the 8 fold Dyson-Altland-Zirnbauer classification of Hamiltonians in the topological phases (see Ref.\cite{stone} for a review). It is natural to expect that certain bulk properties of such systems inherit the similar 8 fold periodicity: see Ref.\cite{DeJonghe:2012rw} for an example. Since we are considering a finite temperature plasma of such particles, we are led to ask a question whether there are characteristic ``hydrodynamics transport properties''  that mirror the underlying classification.
A few simple things can be easily observed. In the momentum flow induced by vorticities only, that is,
\be
T^{0i}\sim \lambda_{(n-1,0)}\epsilon^{0i i_1j_1\cdots i_{n-1}j_{n-1}}(\partial_{i_1}u_{j_1})\cdots (\partial_{i_{n-1}}u_{j_{n-1}})\,,
\ee
the transport coefficient $\lambda_{(n-1,0)}$ is proportional to $B_{n+1}(1/2+\beta\mu/(2\pi i))$. Using the property $B_m(1-x)=(-1)^m B_m(x)$,
this does not vanish in the neutral system ($\mu=0$) only if $n=2k+1$, equivalently in $2n=4k+2$ dimensions. 
This seems to be related to that pure gravitational anomaly exists only in such dimensions.
Similarly, the current induced by vorticities only (whose transport coefficient is $\xi_{(n-1,0)}$) is proportional to $B_{n}(1/2+\beta\mu/(2\pi i))$, which
does not vanish in a neutral system only if $n=2k$, or in $2n=4k$ dimensions. In $2n=8k+2$ dimensions, one can reduce a Weyl spinor further to be Majorana
which violates charge conjugation (C) maximally, and one can't introduce U(1) charge in the system. What would be a characteristic hydrodynamic property
of this system that is distinctive compared to $2n=8k+6$? One promising direction might be to classify the transport coefficients in terms of discrete C, P, T symmetries \cite{Kharzeev:2011ds}.

One may repeat our computations including the damping rate in the propagators. In four dimensions, it has been shown that the damping rate representing a relaxation dynamics due to a finite interaction does not change the CME current \cite{Satow:2014lva}, and we would naturally expect the same in higher dimensions as well. It would be useful to check this explicitly.

Another microscopic framework at weak coupling is the kinetic theory. It would be interesting to check our results in the recently developed chiral kinetic theory \cite{Son:2012wh,Stephanov:2012ki,Gao:2012ix}, suitably generalized to higher dimensions as in Ref.\cite{Dwivedi:2013dea}. We leave this as a future problem.

\vskip 1cm \centerline{\large \bf Acknowledgment} \vskip 0.5cm

 We thank Daisuke Satow for an early collaboration, Tom Imbo and Misha Stephanov for helpful discussions.



\vfil

	\end{document}